\documentclass[twocolumn,showpacs,preprintnumbers,amsmath,amssymb,floatfix]{revtex4}

\usepackage[latin1, ansinew]{inputenc}
\usepackage{graphicx}
\usepackage{dcolumn}
\usepackage{bm}

\begin{document}

\title{Stress-strain behavior and geometrical properties of packings of elongated particles}

\author{Emilien Az\'ema and Farhang Radja\"i}

\affiliation{LMGC, CNRS - Universit\'e Montpellier 2, Place
Eug\`ene Bataillon, 34095 Montpellier cedex 05, France}

\email{azema@lmgc.univ-montp2.fr ; radjai@lmgc.univ-montp2.fr }

\date{\today}

\begin{abstract}

We present a numerical analysis of the effect of particle elongation on  
the quasistatic behavior of sheared granular media by means of  the 
Contact Dynamics method. The particle shapes are rounded-cap 
rectangles characterized by their elongation.      
The macroscopic and microstructural properties of several packings 
subjected to biaxial compression are  analyzed 
as a function of particle elongation. 
We find that the shear strength is an increasing linear function 
of elongation. Performing an additive decomposition of the stress tensor 
based on a harmonic approximation of the angular dependence of 
branch vectors, contact normals and forces, we show that the increasing mobilization of 
friction force and the associated
anisotropy are key effects of particle elongation. 
These effects are correlated with partial nematic ordering of the 
particles which tend to be oriented perpendicular to the major principal stress 
direction and form side-to-side contacts. 
However, the force transmission is found to be mainly guided by 
cap-to-side contacts, which represent the largest fraction of contacts for the 
most elongated particles. Another interesting finding is that, in contrast 
to shear strength, the solid fraction first increases with particle elongation, 
but declines as the particles become more elongated.
It is also remarkable that the coordination number does not follow this 
trend  so that the packings 
of more elongated particles are looser but more strongly connected.

\end{abstract}

\maketitle

\section{Introduction}

Since a few years, the research for a better understanding of the complex rheology of 
granular media is enriched by an increasing focus on nonspherical particles \cite{Richefeu2006,Richefeu2007,
Nouguier-Lehon2003,Antony2004,Kruyt2004,Nouguier-Lebon2005,Alonso-Marroquin2005,Pena2006a,Voivret2009}. 
The wide-spread use of spherical or disk-like particles has been motivated by the 
fact that the rheology of granular media is basically governed by the collective 
contact interactions of the particles so that the particle shape 
can be viewed as a secondary effect. In practice, both in experiments and discrete element 
simulations, the spherical or circular particles, such as glass beads and disks, 
are easier to handle and the 
results are generally more directly amenable to theoretical analysis. However, owing to the fast 
progress in experimental and numerical techniques during the last decade, 
there is now a wide scope for the investigation of materials composed of 
more complex particle shapes. In this respect, the model granular 
media with  spherical particles provide a   
reference material for understanding the 
rheology when the particle shapes deviate from a spherical 
or circular shape \cite{Azema2006,Azema2009}.              

A wide variety of particle shapes can be found in nature and 
industry: elongated and platy shapes, e.g. in biomaterials and pharmaceutical applications,  
angular and facetted shapes, e.g. in geomaterials, 
and nonconvex shapes, e.g. in sintered powders. 
The behavior under various types of loading is strongly influenced by 
particle shape. Rounded particles enhance flowability whereas   
angular shape is susceptible to improve shear strength. In many applications,  
the particle shapes need to be optimized in order 
to increase performance \cite{Markland1981,Wu2000,Lim2005,
Saussine2006,Lobo-Guerrero2006,Lu2007}. 
These trends are generally explained in qualitative terms and linked with 
the jamming of the particles. 

The effect of particle shape is mediated by the specific {\em granular texture}  (or fabric) 
induced by each particle shape.    
For example, it is found that   
hard ellipses can be jammed even though they are underconstrained 
\cite{Tkachenko1999,Donev2004,
Donev2004a,Man2005,Donev2007,Yatsenko2007}. 
In general, the anisometric or elongated particle shapes, such as spheroids and sphero-cylinders,  
tend to develop orientational order 
affecting force transmission and frictional 
behavior \cite{Ouadfel2001,Nouguier-Lehon2003,Zuriguel2007,Hidalgo2009}. 
This ``nematic'' ordering occurs while, in contrast to liquid crystals, 
the particles interact only via contact and friction \cite{Kyrylyuk2009}.   

In a sheared granular material, the local equilibrium structures are generically 
anisotropic in terms of contact directions and forces 
\cite{Kruyt1996,Bathurst1988,Rothenburg1989,Radjai1998,mirghasemi2002,Troadec2002,Kruyt2004,Radjai2009a}. 
It was recently shown that the 
fabric anisotropy in a sheared granular assembly crucially depends on 
particle shape \cite{Azema2006,Azema2009}. 
In the case 
of polygonal and polyhedral particles, due to large contact 
area  of side-to-side contacts, the fabric
anisotropy appears to be marginal  compared to force anisotropy \cite{Azema2006,Azema2009}.
Those contacts play a major role in force
transmission by accommodating long force chains that are basically 
unstable in a packing composed of spheres. 

The force and fabric anisotropies are at the origin of the 
enhanced shear strength of materials composed of nonspherical particles
\cite{Azema2007,Azema2009,Ouadfel2001,mirghasemi2002}. 
The particle shape affects the compactness and dilatancy of granular materials.
A nontrivial effect, evidenced recently by experiments and numerical simulations for 
spheroids, is the finding that the solid fraction is not a monotonous function of the 
aspect ratio \cite{Donev2004,Donev2004a,Man2005,Donev2007,Sacanna2007}. The solid fraction increases linearly to a maximum 
and then declines in inverse proportion to the aspect ratio  \cite{Williams2003}.  
In powder processing, the particle shape appears also to be an important parameter 
controlling the flowability, discharge rates and compaction of 
powders \cite{fraige2008,Langston2004}. 

In this paper, we use contact dynamics simulations to investigate 
the rheology of large packings of elongated particles with increasing 
aspect ratio. The particles are rectangles with rounded caps to which we will refer as 
Rounded-Cap Rectanglular (RCR) particles. 
These particles may be 
considered as 2D analog of sphero-cynlinders. The RCR shape can be characterized by 
a single aspect ratio $\alpha$ or, as we shall see, by an elongation parameter $\eta$ varying 
from $0$ to $1$ as the particle shape varies continuously from a circle to an 
thin line. We are interested both in the properties of the static packings 
of RCR particles prepared by isotropic compaction without friction 
and in the stress-strain behavior under 
biaxial compression with finite friction between particles. 

The macroscopic behavior is studied in terms of the internal angle of friction 
and solid fraction for different values of $\eta$. We find a nonmonotonous 
variation of the solid fraction and a nearly linear increase of the internal angle of friction 
with $\eta$. In order to understand the origins of this behavior and the role  
of particle shape, we perform a detailed analysis of 
the microstructure and stress transmission.  
We consider the organization of the particles and contacts 
in the simulated packings, as well as the  
stress transmission by means  of a harmonic representation 
of the stress tensor in terms of force and fabric anisotropies.  
The microstructure is increasingly dominated by a short-range nematic ordering of particle 
orientations as $\eta$ increases. 
We show that the internal angle of friction is influenced by this ordering via 
an increasing anisotropy of friction forces and contact orientations with 
the elongation parameter. For all values of the latter, the harmonic 
approximation provides an excellent fit to the shear stress. 

An important feature of RCR particles is that, like polygonal particles, 
they have lineal edges and can thus form side-to-side contacts 
as well as side-to-cap and cap-to-cap contacts. Hence, in a packing of RCR 
particles, the texture can characterized by the networks of these various 
contact types, and the influence of the shape parameter on force transmission 
and shear strength may be analyzed in terms of these contacts and more specially 
the side-to-side contacts which are expected to play a stabilizing role in the packing.

In the following, we first introduce our numerical approach in Section \ref{procedure}. Then, in 
Section \ref{Strength}, the stress-strain behavior is presented for different values of $\eta$.  
The microstructure is analyzed in Section \ref{harmonics} in terms of connectivity, orientations of the 
particles and the contact network. We also introduce the
 harmonic approximation of the stress tensor allowing us to track the 
 origins of the internal angle of friction via force and fabric anisotropies. 
In Section \ref{Geom_mec_origins}, we present an additive  decomposition of the connectivity, 
anisotropies and forces as a function of different contact types.  
The force distributions are presented in Section \ref{distribution}. In section 
\ref{contact_type}, we analyze the structure of force networks with cap-to-cap, cap-to-side and side-to-side
contacts. We conclude 
with a summary and discussion of the most salient results of this work.

\section{System description and numerical procedures}
\label{procedure}
In this section, we briefly introduce the contact dynamics (CD) method used for 
the simulations and the numerical procedures for sample preparation.   

\subsection{Contact Dynamic method}
The simulations were carried out by means of the contact dynamics (CD) 
method \cite{Jean92,Moreau1994,Jean1999,Moreau2004,Radjai1999,Dubois2003,Renouf2005,Richefeu2007,
Radjai1997,Radjai2009}. 
The CD method is a discrete element approach based on a nonsmooth approach 
in which an integrated form of the equations of dynamics. The integration 
time interval corresponds to the time step and may involve discontinuous 
variation of the velocities due to collisions.         
The frictional and collisional interactions are described 
as {\it complementarity relations} between 
the relative velocities between particles and the corresponding momenta 
at the contact points without elastic or viscous regularization.
Thus, the condition of geometrical contact  between two  particles is expressed by  
the following mutually exclusive alternatives:
\begin{equation}
\begin{array}{lll}
f_n \geqslant 0 & \mbox{and}& u_n = 0,\\
f_n = 0              & \mbox{and}& u_n > 0.
\end{array}
\label{eq:1}
\end{equation}
where $f_n$ is the normal contact force and $u_n$ 
the relative normal velocity between two particles in contact is counted 
positive when they move away from each other.

In the same way, the Coulomb friction law involves three mutually exclusive conditions:
\begin{equation}
\begin{array}{cll}
f_t = -\mu f_n                                         & \mbox{and}&       u_t > 0, \\
-\mu f_n \leqslant f_t \leqslant \mu f_n & \mbox{and}&       u_t = 0, \\
f_t = \mu f_n                                         & \mbox{and}&       u_t < 0, \\
\end{array}
\label{eq:2}
\end{equation}
where $u_t$ is the sliding velocity at the contact, $\mu$ is the friction 
coefficient and $f_t$ is the friction force.
Remark that this 
relation cannot be reduced to a (mono)valued functional dependence 
between the two variables as assumed in the Molecular Dynamics (MD) method.

The above formulation is implicit  in the sense that 
the complementarity relations should be satisfied for the velocities at 
the end of each time step.  An iterative algorithm based on a  nonlinear 
Gauss-Seidel scheme is used to 
solve the system of equations and complementarity relations for contact 
forces and particles velocities.  
The uniqueness is not {\em a priori} guaranteed for perfectly rigid particles. 
However, by initializing each step of calculation with the forces calculated  in the 
preceding step, the set of admissible solutions shrinks to fluctuations which are basically 
below the numerical resolution.  

The CD method is particularly suitable for the simulation of rigid nearly undeformable 
particles. In this limit, the MD method requires steep interaction potentials 
and thus very small time steps. Nevertheless, several comparisons between 
the two methods suggest that both methods are equally valid and efficient for 
the simulation of granular materials \cite{Richefeu2007,Radjai1997,F.Radjai2009}.       

\subsection{Simulation of RCR particles}

\begin{figure}
\includegraphics[width=4cm]{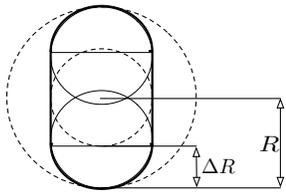}
\caption{Shape of a Rounded-Cap Rectangle (RCR). 
 \label{sec:numerical_procedure:def_jonc}}
\end{figure}

We model the RCR particle as a juxtaposition of two half-disks of radius $R'$ 
with one rectangle of length $L$ and width $2R'$; see 
Fig. \ref{sec:numerical_procedure:def_jonc}. 
The shape of a RCR particle is a circle of radius $R'$ for $L=0$. The aspect ratio 
$\alpha = (L+2R')/(2R')$  is $1$ in this limit and increases with $L$ for a fixed value of $R'$.  
In this paper, we use an alternative parameter describing the deviation of the particle shape 
from a circle. Let $R$ be the radius of the circle circumscribing the particle. 
We have $R=L/2+R'$. The radius $R'$ is also that of the inscribed circle. 
Hence, the deviation from a circular shape can be characterized by $\Delta R = R-R' = L/2$. 
We use the dimensionless parameter $\eta$ defined by 
\begin{equation}
\eta = \frac{\Delta R}{R} = \frac{\alpha-1}{\alpha}.
\end{equation}
It varies from $\eta=0$, for a circle,  
to 1 corresponding to a line. We will refer to $\eta$ as the {\em elongation} 
parameter as in rock mechanics \cite{Folk1974}. 

\begin{figure}
\includegraphics[width=2.5cm]{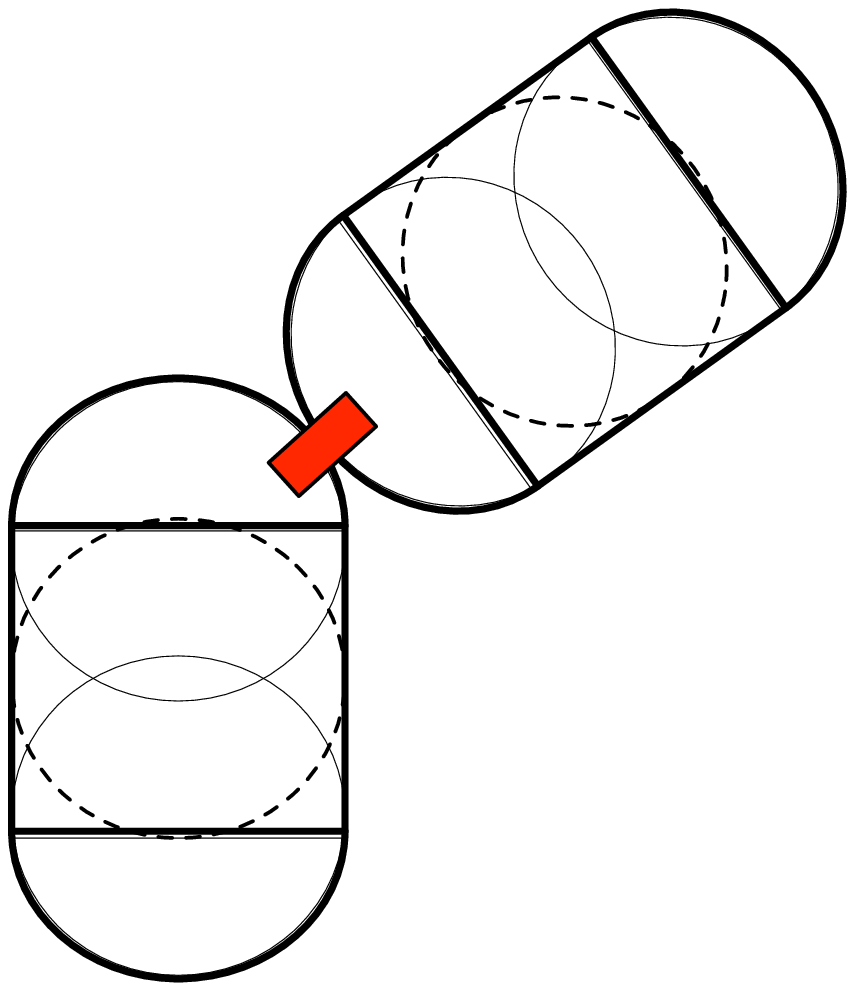}
\includegraphics[width=2.5cm]{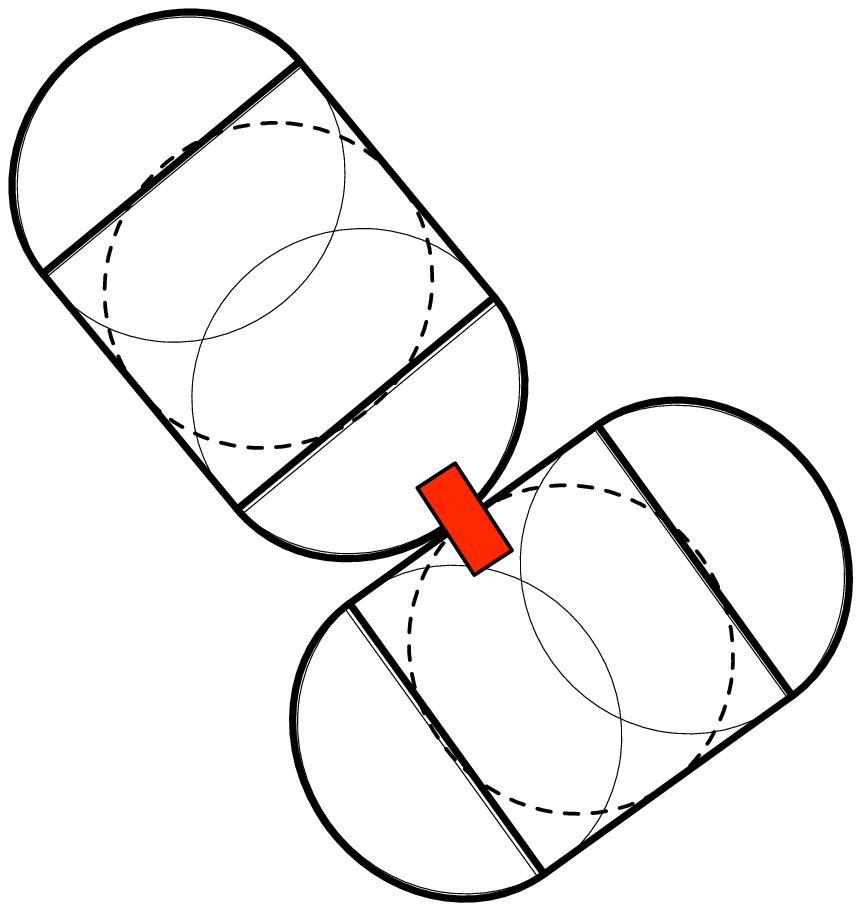}
\includegraphics[width=2.5cm]{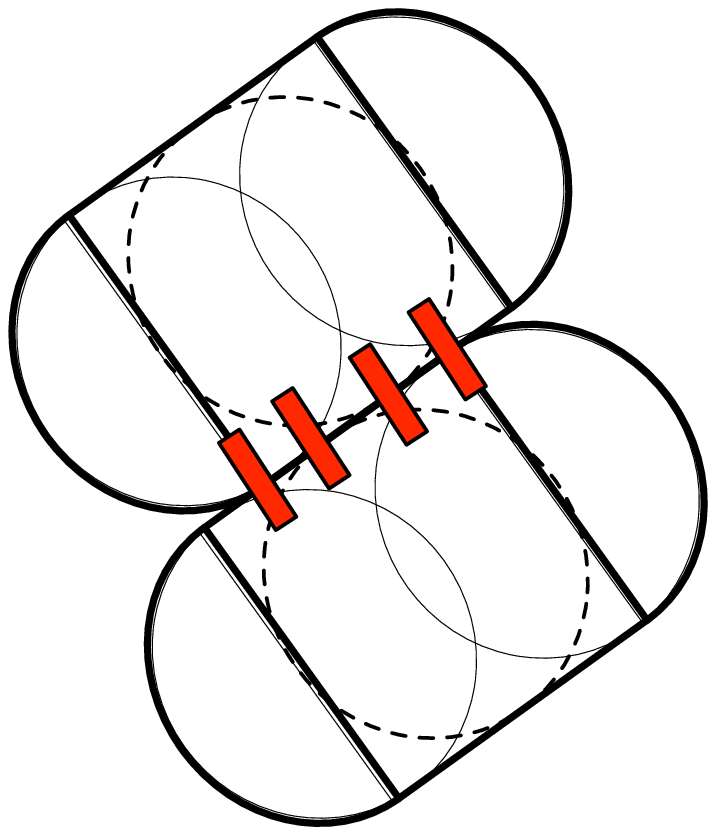}
\caption{Representation of cap-to-cap, cap-to-side  
 and side-to-side contact and they will be referred  as  $cc$ contacts, $cs$ contacts and $ss$ contact, respectiveley.\label{sec:numerical_procedure:type_contacts}}
\end{figure}

For the detection of the contacts between two RCR particles, we use 
the schema shown in  Fig. \ref{sec:numerical_procedure:type_contacts}. 
Three types of contact can be distinguished: cap-to-cap ($cc$), cap-to-side ($cs$) and 
side-to-side ($ss$). The contacts between the particles are thus detected separately for 
the pairs of circles and rectangles. In general, in the CD method $ss$ contact between 
two rectangles   
is treated as composed of two point contacts and the contact laws  (\ref{eq:1}) and  (\ref{eq:2})  
are applied separately to each point. The choice of these points does not affect 
the resultant force and its point of application.  Hence, for RCR particles, as 
shown in Fig. \ref{sec:numerical_procedure:type_contacts}, $ss$ contact is 
composed of four point contacts : two points due to the 
rectangle-rectange interface and two points due to the $cc$ contacts. 
Thus, four forces are calculated by the CD algorithm but only their 
resultant and application point are material. 

The detection of line contacts between rectangles was implemented through 
the so-called {\it shadow overlap method} devised initially by 
Moreau \cite{Dubois2003,Saussine2006} for polygons. The reliability and robustness of this method 
have been tested in several 
years of previous applications to  granular materials 
\cite{Nouguier-Lehon2003,Cholet2003,Azema2006,Saussine2006,Azema2007,Azema2008,Azema2009}. 
This detection procedure is fairly rapid and allows us to simulate large samples 
composed of RCR particles. 
For our simulations, we used the $\bm{LMGC90}$ which is a multipurpose 
software developed in 
Montpellier, capable of modeling a collection of deformable or undeformable particles of 
various shapes (spherical, polyhedral, or polygonal) by means of the 
contact dynamics (CD) method
\cite{Dubois2003}.

\subsection{Sample preparation}
We prepared 8 different packings of $13000$ RCR particles 
with $\eta$ varying  from 0 to $0.7$ by steps of $0.1$. 
The radius $R$ of the 
circumscribing circle defines the size of a RCR particle. 
In order to avoid long-range ordering in the limit of small values 
of $\eta$, we introduce a size polydispersity by taking $R$ in 
the range $[R_{min} , R_{max} ]$ with $R_{max} = 3R_{min}$ 
with a uniform distribution in particle volume fractions.

All samples are prepared according to the same protocol. A dense 
packing composed of disks ($\eta=0$) is first constructed by means of a 
layer-by-layer deposition model based on simple geometrical rules 
\cite{Bratberg2002,Taboada2005,Voivret2007,Voivret2008}. 
The particles are deposited sequentially on a substrate. Each new particle
is placed at the lowest possible position at the free surface as a 
function of its diameter. This procedure leads to a random close packing
in which each particle is supported by two underlying particles and
supports one or two other particles. For $\eta > 0$, the same packing is used 
with each disk serving as the circumscribing circle of a RCR particle.   
The RCR particle is inscribed with the given value of $\eta$ and random orientation 
in the disk.

\begin{figure}
\includegraphics[width=8.5cm]{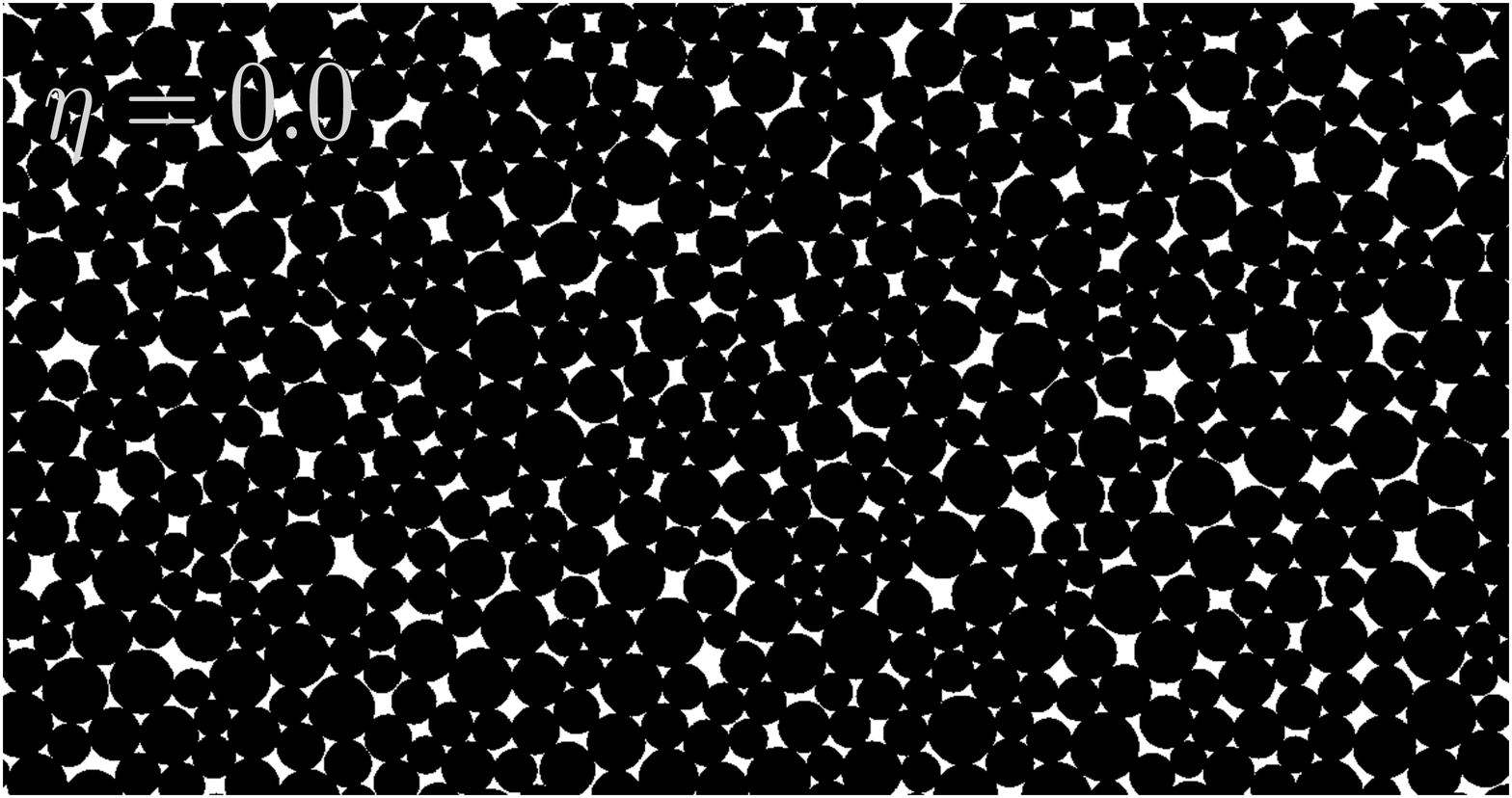}
\includegraphics[width=8.5cm]{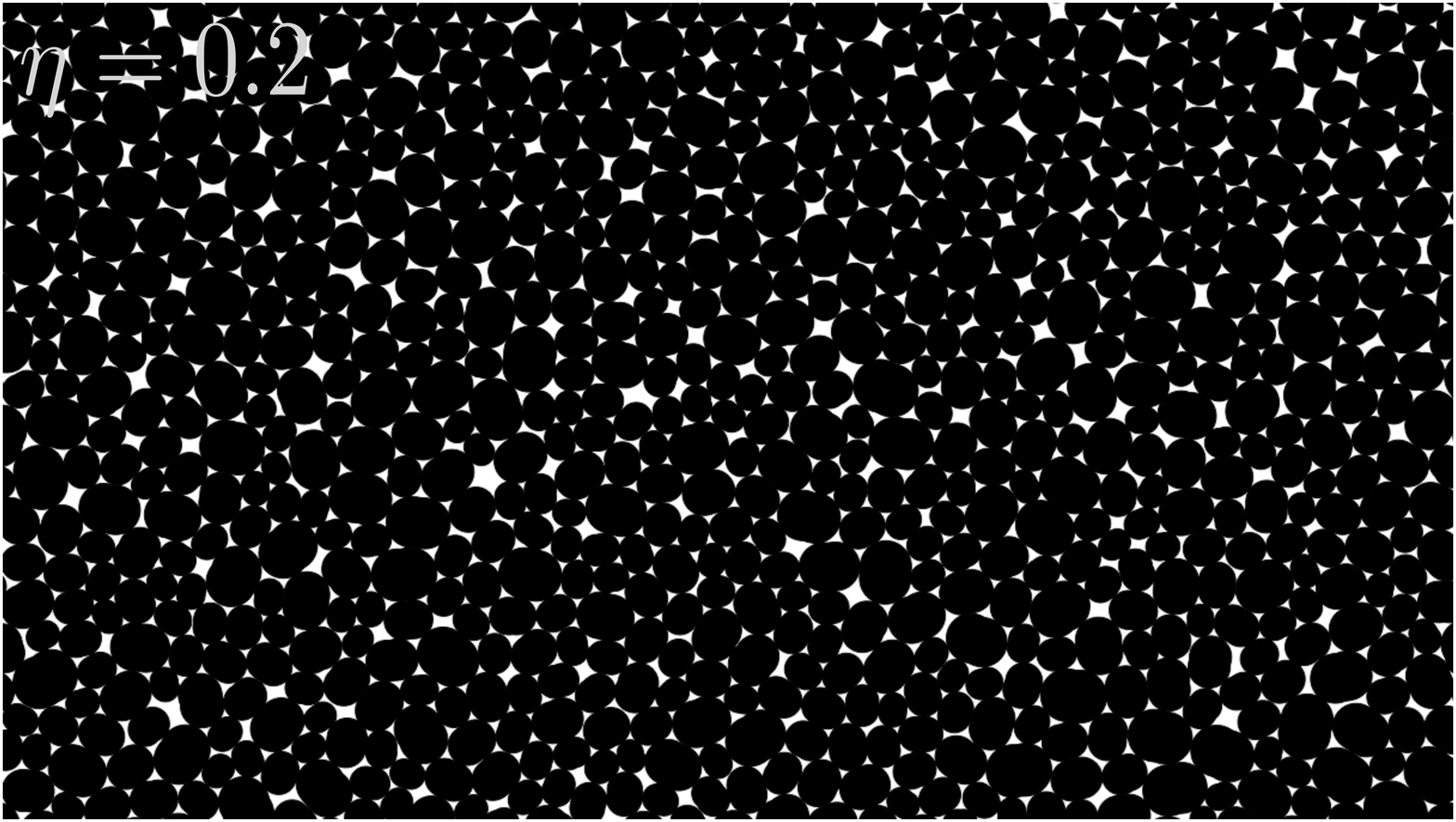}
\includegraphics[width=8.5cm]{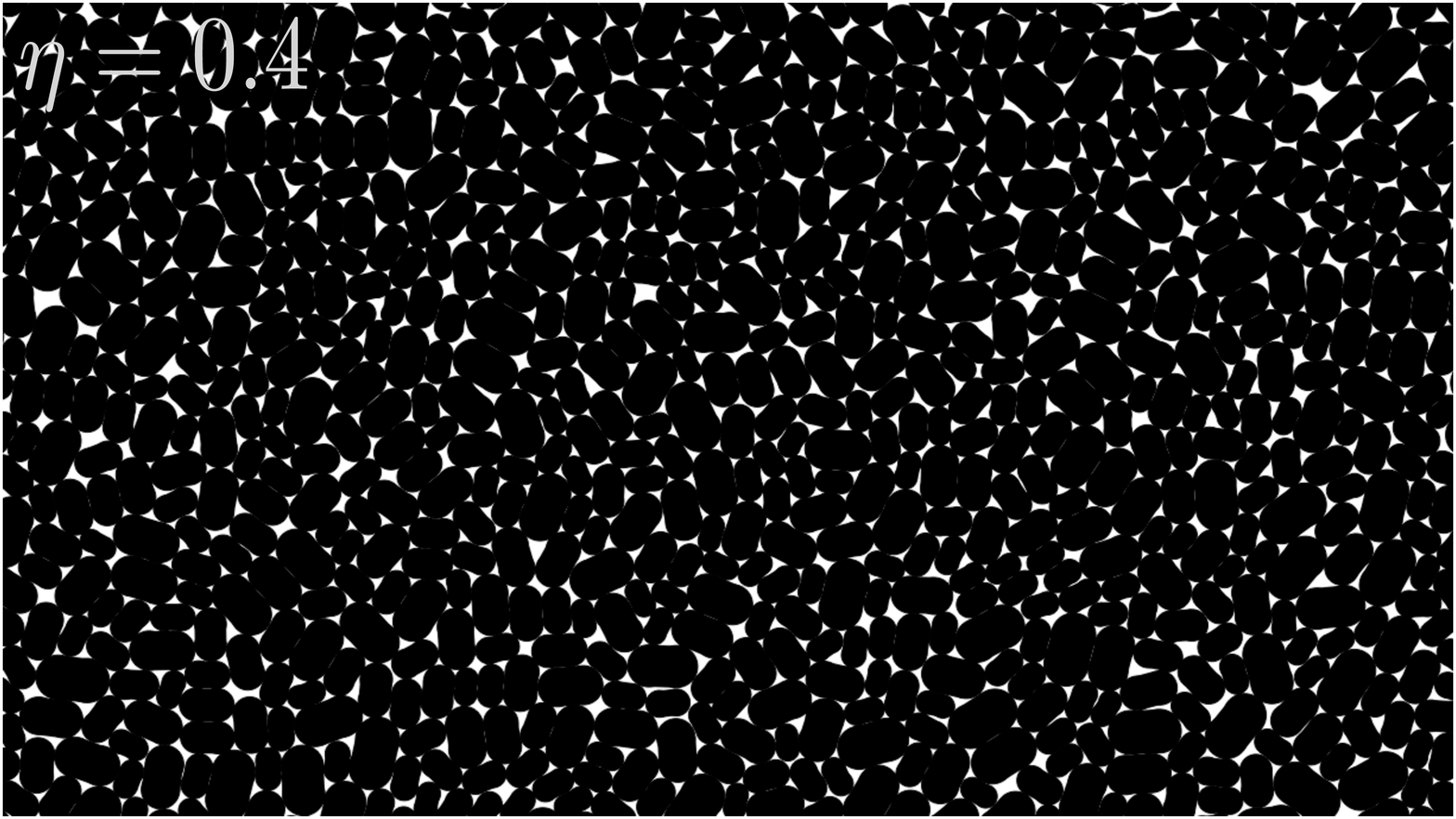}
\includegraphics[width=8.5cm]{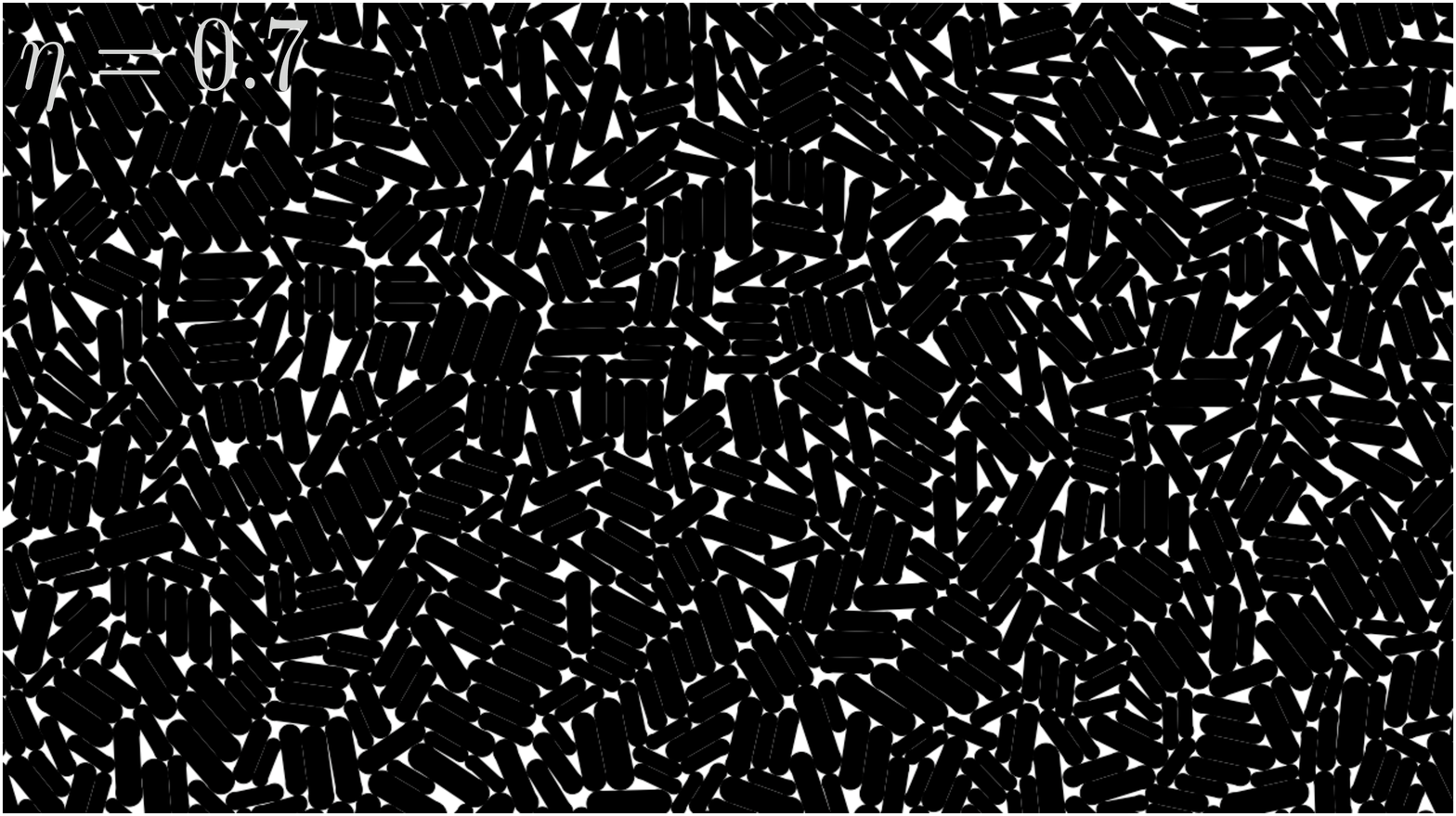}
\caption{Examples of the generated packings at the initial state.  \label{map_ini}}
\end{figure}

Following this geometrical process, the packing is compacted by isotropic 
compression inside a rectangular frame of dimensions $l_0 \times h_0$ in which 
the left and bottom walls are fixed, and the right and top walls are subjected to 
a compressive stress $\sigma_0$. 
The gravity $g$ and friction coefficients $\mu$  between particles 
and with the walls are set to zero during the compression in order to
avoid force gradients and obtain isotropic dense packings. 
Fig. \ref{map_ini}  displays snapshots of the packings for 
several values of $\eta$ at the end of isotropic compaction. 

The isotropic samples are then subjected to vertical compression by downward 
displacement of the top wall at a constant velocity $v_y$ for 
a constant confining stress $\sigma_0$ acting on the lateral walls.
The friction coefficient $\mu$  between particles is set to $0.5$ and 
to zero with the walls.
The simulations were run with a time step 
of   $2.10^{-4}$ s.  The CPU time was  $5.10^{-4}$ s
per particle and per time step on an AMD processor.  
Since we are interested in quasistatic behavior, the shear rate should be 
such that the kinetic energy supplied by shearing is negligible compared to 
the static pressure. 
This can be formulated in terms of an {\it inertia parameter} $I$  defined by \cite{GDR-MiDi2004}:
\begin{equation}
I=\dot \varepsilon \sqrt{\frac{m}{p}},
\label{eq3}
\end{equation}
where $\dot \varepsilon=\dot y /y$ is the strain rate,  
$m$ is the particle mass and $p$ is the mean pressure. The quasistatic limit is
characterized by the condition $I\ll1$. In our simulations, $I$ was 
below $10^{-3}$.  

\section{Strength and dilatancy}
\label{Strength}

In this section, we consider the stress-strain and volume-change behavior as a function 
of the shape parameter $\eta$. We need to evaluate the stress 
tensor and solid fraction  during 
deformation from the simulation data. For the stress tensor, we start with  the tensorial 
moment ${\bm M}^i$ of each particle i that is  defined by \cite{Moreau1997,Staron2005}:
\begin{equation}
M^i_{\alpha \beta} = \sum_{c \in i} f_{\alpha}^c r_{\beta}^c,
\label{eq:M}
\end{equation}
where  $f_{\alpha}^c$ is the $\alpha$ component of the force exerted on 
particle i at the contact c, $r_{\beta}^c$ is the $\beta$ component 
of the position vector of the same contact c, and the summation 
runs over all contact neighbors of particle i (noted briefly by $c \in i$).
The average stress tensor $\bm \sigma $ in the volume $V$ of the granular 
assembly is given by the sum of the tensorial moments of 
individual particles divided by the volume \cite{Moreau1997,Staron2005}:  
\begin{equation}
{\bm \sigma } = \frac{1}{V} \sum_{i \in V} {\bm M}^i =   \frac{1}{V}  \sum_{c \in V} f_{\alpha}^c \ell_{\beta}^c, 
\label{eq:sigma}
\end{equation}
where ${\bm \ell}^c$ is the branch vector joining the centers of 
the two touching particles at the 
contact point $c$. Remark that the first summation runs over all 
particles whereas the second summation 
involves the contacts, each contact appearing only once.  

Under biaxial conditions with vertical compression, we have 
$\sigma_1 \geq \sigma_2$, where the $\sigma_\alpha$ are 
the stress principal values.  
The mean stress $p$ and stress deviator $q$ are defined by: 
\begin{eqnarray}
p &=&  \frac{1}{2}(\sigma_1 + \sigma_2), \label{eq:p} \\
q &=& \frac{1}{2}(\sigma_1 - \sigma_2). \label{eq:q}
\end{eqnarray}
For our system of perfectly rigid particles,  the stress state is characterized by 
the mean stress $p$ and normalized shear stress $q/p$.    

The strain parameters are the cumulative vertical, horizontal and 
shear strains $\varepsilon_1$, $\varepsilon_2$ and $\varepsilon_q$, respectively.
By definition, we have 
\begin{equation}
\varepsilon_1=\int_{h_0}^h \frac{dh'}{h'} = \ln \left( 1+ \frac{\Delta h}{h_0} \right),
\end{equation}
where $h_0$ is the initial height and $\Delta h = h_0 - h$ is the total downward displacement, and 
\begin{equation}
\varepsilon_2=\int_{l_0}^l \frac{dl'}{l'} = \ln \left( 1+ \frac{\Delta l}{l_0} \right),
\end{equation}
where $l_0$ is the initial box width and $\Delta l = l - l_0$ is the total change of the box width. The cumulative shear strain is then defined by
\begin{equation}
\varepsilon_q \equiv \varepsilon_1-\varepsilon_2.
\label{eq:qV}
\end{equation}
Finally, the cumulative volumetric strain $\varepsilon_p$  is given by
\begin{equation}
\varepsilon_p=\varepsilon_1 + \varepsilon_2 =\int_{V_0}^V \frac{dV'}{V'} = \ln \left( 1+ \frac{\Delta \nu}{\nu} \right)
\end{equation}
where $V_0=l_0 h_0$ is the initial volume 
and $\Delta \nu = \nu - \nu_0$ is the cumulative change of solid fraction.

\begin{figure}
\includegraphics[width=8cm]{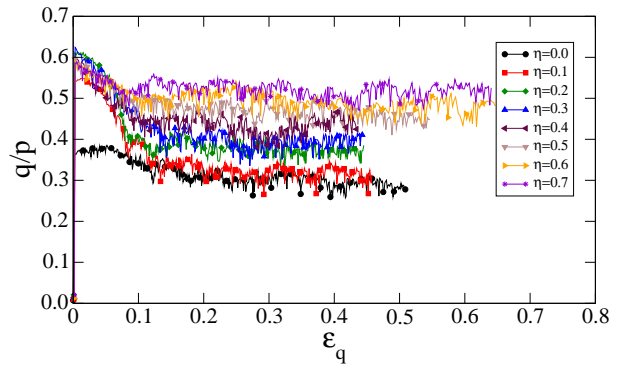}
\caption{Normalized shear stress $q/p$  as a function of
cumulative shear strain $\varepsilon_q$ for different values of 
the shape parameter $\eta$.}  
\label{sec:stress:qsurp_ep_eq}
\end{figure}

Figure \ref{sec:stress:qsurp_ep_eq}  shows the normalized shear stress $q/p$  
as a function of shear strain $\varepsilon_q$ for different values of $\eta$.       
The jump observed at $\varepsilon_q=0$ reflects both the rigidity of the particles and 
high initial solid fraction of the samples (see below). In all cases, the shear stress passes by a peak before relaxing 
to a stress plateau corresponding to the so-called ``residual state'' 
in soil mechanics \cite{Mitchell2005}. We remark that the residual shear stress 
increases with $\eta$. 

The internal angle of friction $\varphi^*$, representing the shear 
strength of the material, is defined from the mean 
value $(q/p)^*$ of the normalized shear stress in the
residual state by \cite{Mitchell2005}
\begin{equation}
\sin \varphi^* = \Big(\frac{q}{p} \Big)^*. 
\end{equation}
Fig. \ref{sec:stress:qsurp_ep_eq} shows the variation of 
$\sin \varphi^*$ as a function of $\alpha$ and $\eta$. 
We see that the shear strength is an increasing nonlinear function of 
the aspect ratio, but, interestingly, it varies linearly when plotted 
versus the elongation parameter. Hence, we have 
\begin{equation}
\sin \varphi^* = \sin \varphi^*_0 + k \ \eta = \sin \varphi^*_0 + k \ \left( 1 - \frac{1}{\alpha} \right) 
\end{equation}
This observation indicates that the evolution of shear strength reflects 
more directly shape elongation than aspect ratio. In the following, we will 
use $\eta$ as shape parameter.

\begin{figure}
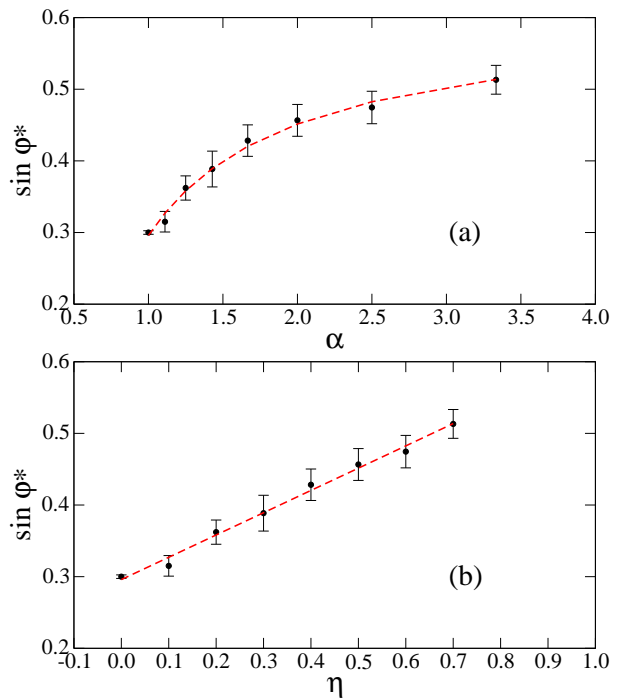

\includegraphics[width=8cm]{fig05a.eps} 
\includegraphics[width=8cm]{fig05b.eps} 
\caption{Internal angle of friction $\varphi^*$ as a function of aspect ratio $\alpha$ (a) 
and elongation $\eta$. The error bars represent the standard deviation 
in the residual state.}
\label{fig:sinphi}
\end{figure}

\begin{figure}
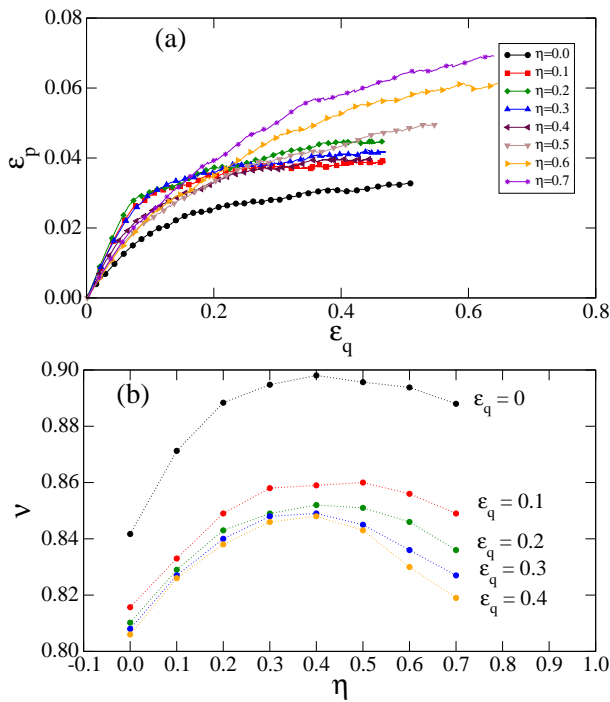

\includegraphics[width=8cm]{fig06a.eps}
\includegraphics[width=8cm]{fig06b.eps}
\caption{Cumulative volumetric strain  
$\varepsilon_p$  as a function of shear strain $\varepsilon_q$ (a);  
Solid fraction as a function of particle shape parameter $\eta$ (b) at 
different levels of shear strain. }  
\label{sec:stress:ep_eq}
\end{figure}

Figure \ref{sec:stress:ep_eq} (a) displays the cumulative volumetric strain  
$\varepsilon_p$ as a function of $\varepsilon_q$ for different values of $\eta$.
Starting with an initially  dense state, all packings dilate and hence the 
volume increases. For $\eta \leq 0.4$, a plateau is reached  
beyond $\varepsilon_q=0.3$, corresponding to a state of isochoric deformation. 
For larger aspect ratios, the dilatation continues even at very large 
deformations. This is an indication of an inhomogeneous dilation due to the 
formation of shear bands in the bulk, which is enhanced by the elongated shape  
of the particles. Since different parts of the packing undergo differential volume change, 
longer shearing is required to reach a fully dilated state for the whole packing. 
The initiation and  evolution of 
shear bands for different particle elongations will be reported in more detail 
elsewhere. 

Figure \ref{sec:stress:ep_eq} (b) displays the solid fraction 
$\nu$ as a function of $\eta$ at different levels of shear 
deformation $\varepsilon_q$. It is remarkable that, at all levels of deformation, 
the solid fraction increases with $\eta$, reaches a maximum 
at $\eta \simeq 0.4$ and then declines as $\eta$ further increases. 
We note that solid fractions as large as  $0.90$
are reached for $\eta=0.4$ in the initial state. 
A similar nonmonotonous behavior was observed for packings of ellipses or 
ellipsoidal particles \cite{Donev2004,Donev2004a,Donev2007}. 
This is somewhat a counterintuitive finding as  
the shear strength (a monotonous function of $\eta$) does not follow 
the trend of solid fraction (nonmonotonous).  
This behavior is clearly not 
related to shear localization since it is observed at all levels of deformation including 
the initial isotropic state  $\nu_0 = \nu(\varepsilon_q=0)$ where the packings 
are homogeneous. 

A rapid fall-off of solid fraction for elongated particles in 3D 
was observed at very large aspect ratios and  attributed to the 
excluded volume due to disorder, predicting  a fall-off in inverse proportion to 
the aspect ratio \cite{Blouwolff2006,Williams2003,Wouterse2007}. 
The initial rapid increase of solid fraction, as observed in 
Fig. \ref{sec:stress:ep_eq}, reveals that excluded-volume effects  
are not the prevailing mechanism at low aspect ratios. In this limit, 
slight deviations from spherical shape have strong space-filling effect 
on the packing although the excluded volume increases at the same 
time and becomes dominant at very large aspect ratios.        

\begin{figure}
\includegraphics[width=8cm]{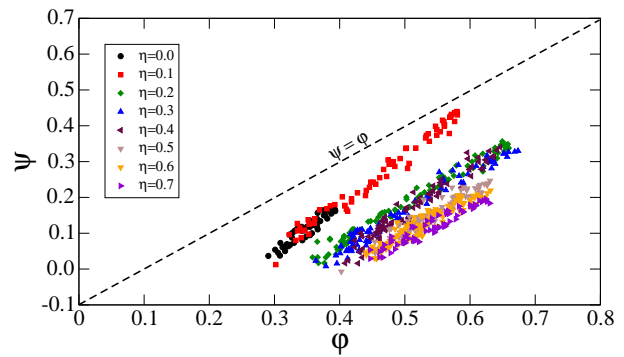}
\caption{Dilatancy angle $\psi$ versus internal angle of friction $\varphi$ 
for different values of $\eta$. 
\label{sec:stress:phi_psi}}
\end{figure}

The volumetric deformation can also be expressed in terms 
of the {\it dilatancy angle} $\psi$ defined by \cite{Wood1990}:
\begin{equation}
\sin \psi = \frac{\partial\varepsilon_p}{\partial\varepsilon_q}. 
\end{equation}
The plot of $\psi$ as a function of $\varphi$, the so-called  ``stress-dilatancy diagram'',  
is shown in Fig. \ref{sec:stress:phi_psi} for different values of $\eta$. 
We observe a linear correlation between $\varphi$ and $\psi$ irrespective 
of the value of $\eta$. We have   
\begin{equation}
\varphi \simeq \varphi^* + \psi. 
\label{eqn:stressdilatancy}
\end{equation}
This is a particularly simple relation compared to several models 
proposed in soil mechanics \cite{Wood1990,Radjai2004}. 
It reflects the ``non-associated'' character of granular plasticity, an associated 
behavior implying simply   $\varphi = \psi$, which is unrealistic for granular 
materials \cite{TAYLOR_1948,Vermeer_1984a,vermeer_1984b,Wood1990}. 
According to relation (\ref{eqn:stressdilatancy}), the dilatancy angle 
vanishes in the residual state where $\varphi = \varphi^*$.   
Recent work on cohesive and granular packings of polygonal particles  in 2D 
is in agreement with this correlation 
\cite{Taboada2006,Azema2007}. 

In the following, we focus on the microstructure and force transmission 
that provide a key to a better understanding of the physical mechanisms underlying the 
effect of particle shape on the shear strength.

\section{Granular textures}
\label{harmonics}

In this section, we investigate the general organization (texture) of our 
packings of RCR particles in terms of particle orientations and 
contact network. This will allow us to quantify the effect of the elongation 
parameter and its connection with shear strength.    

\subsection{Particle orientations}
\begin{figure}
\includegraphics[width=8cm]{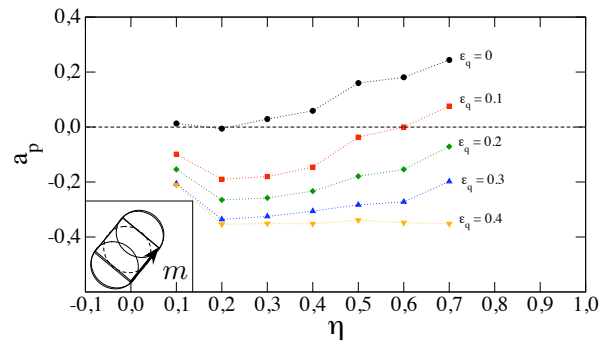}
\caption{Particle orientation anisotropy $a_p$ as a function of 
shape parameter $\eta$ at different stages of shear $\varepsilon_q$.}
\label{sec:harmonic:orientation_part}
\end{figure}

The principal feature of elongated particles is their orientational degree 
of freedom.   The particle orientation is represented by a  
unit vector $\bm m$ as shown in the inset to Fig. \ref{sec:harmonic:orientation_part}. 
In 2D, it is parametrized by a single angle $\vartheta$. Let $\cal D(\vartheta)$ be the set of
particles with direction 
$\vartheta \in[\vartheta - \delta \vartheta/2 ; \vartheta + \delta \vartheta / 2]$ for angle
increments $\delta \vartheta$, and $N_p(\vartheta)$ its cardinal.
The probability $P_\vartheta (\vartheta)$ of 
the orientations of particles is given by 
\begin{equation}
P_\vartheta(\vartheta) = \frac{N_p(\vartheta)}{N_p},
\end{equation}
where $N_p$ is the total number of particles.

\begin{figure}
\includegraphics[width=8cm]{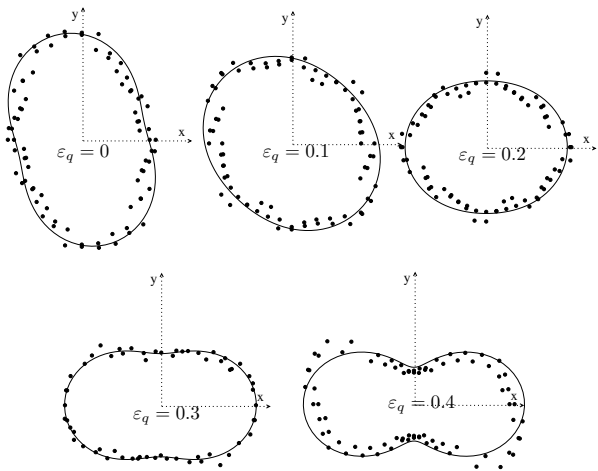}
\caption{Polar representation of the probability density $P_\vartheta$ 
of  particle orientations $\vartheta$ for
$\eta=0.7$ at various stages of deformation $\varepsilon_q$.
The symbols are the simulations data. Solid lines represent harmonic fit according 
to equation (\ref{fit_ap}).
\label{sec:harmonic:orientation_grain}}
\end{figure}

Figure \ref{sec:harmonic:orientation_grain} displays a polar representation of 
$P_\vartheta(\vartheta)$ for $\eta=0.7$ at various stages of deformation. 
In the initial state, corresponding to an isotropic stress state, the 
distribution is anisotropic with privileged direction $\vartheta_p$ 
close to  $\pi/2$. This means that for elongated particles, 
the particle orientations are not fully correlated with the stress state 
so that the resulting particle orientation anisotropy depends on 
details of the assembling process that can not be controlled by 
simply subjecting the particles to isotropic stresses from the boundary.          

The priviliged direction rotates as a result of vertical 
compression and becomes horizontal (parallel to the minor principal stress direction) 
in the residual state. The distribution are nicely fitted by harmonic approximation  
corresponding to the lowest order terms of the Fourier expansion of  
$P_\vartheta(\vartheta)$ \cite{Rothenburg1993,Reddy2009}:
\begin{equation}
P_\vartheta(\vartheta) = \frac{1}{2\pi} (1 + a_p\cos (2(\vartheta - \theta_\sigma)),
\label{fit_ap}
\end{equation}
where $a_p$ represents the anisotropy of the distribution and $\theta_\sigma$ is 
the major principal stress direction. The choice of  $\theta_\sigma$ as reference 
direction is motivated by the observation that the privileged orientation of the 
particles tends to align itself with the minor principal stress direction. Hence, 
negative values of $a_p$ mean that the particles are preferentially  
oriented perpendicular to the major principal stress. 

\begin{figure}
\includegraphics[width=9cm]{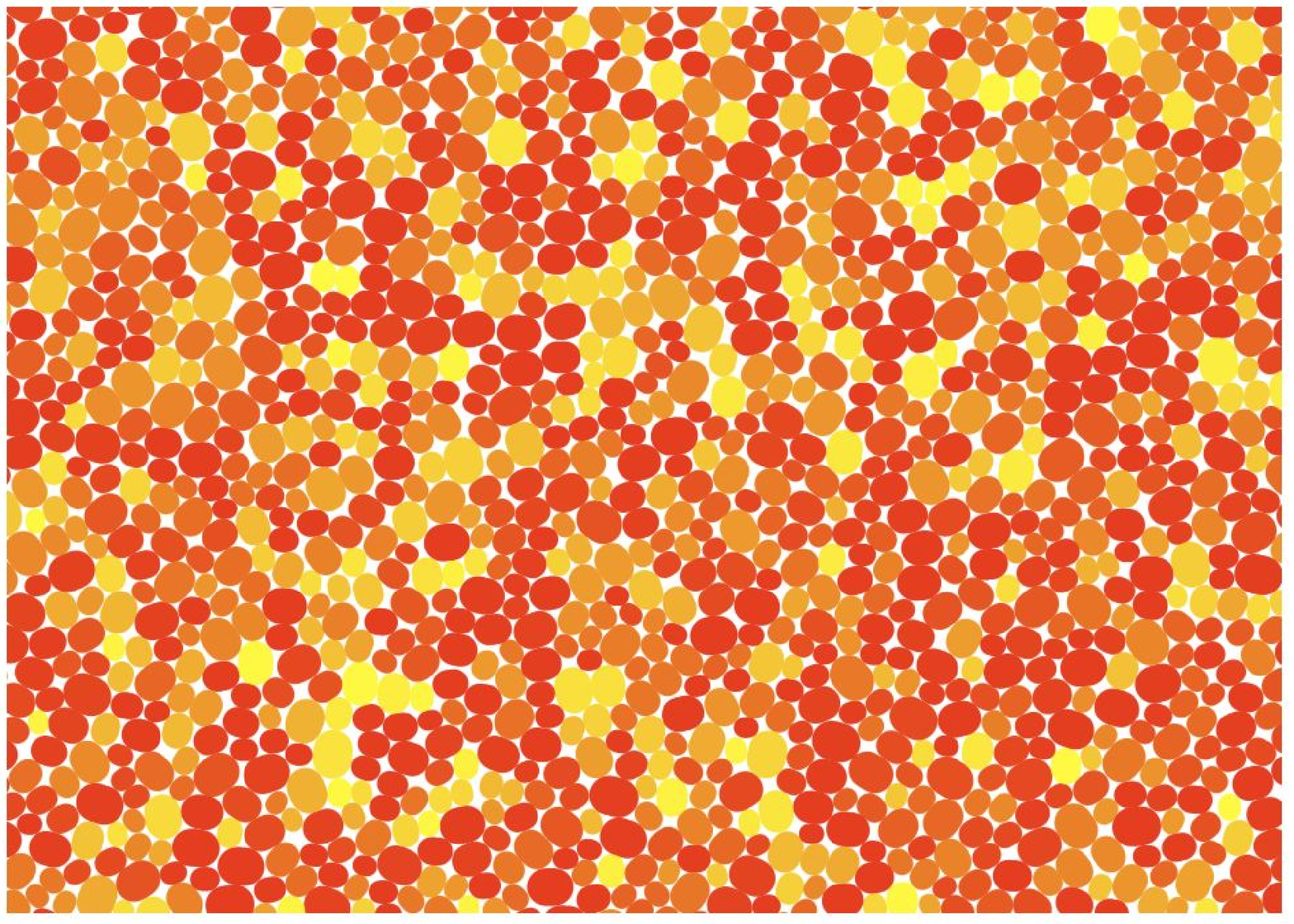}
\includegraphics[width=9cm]{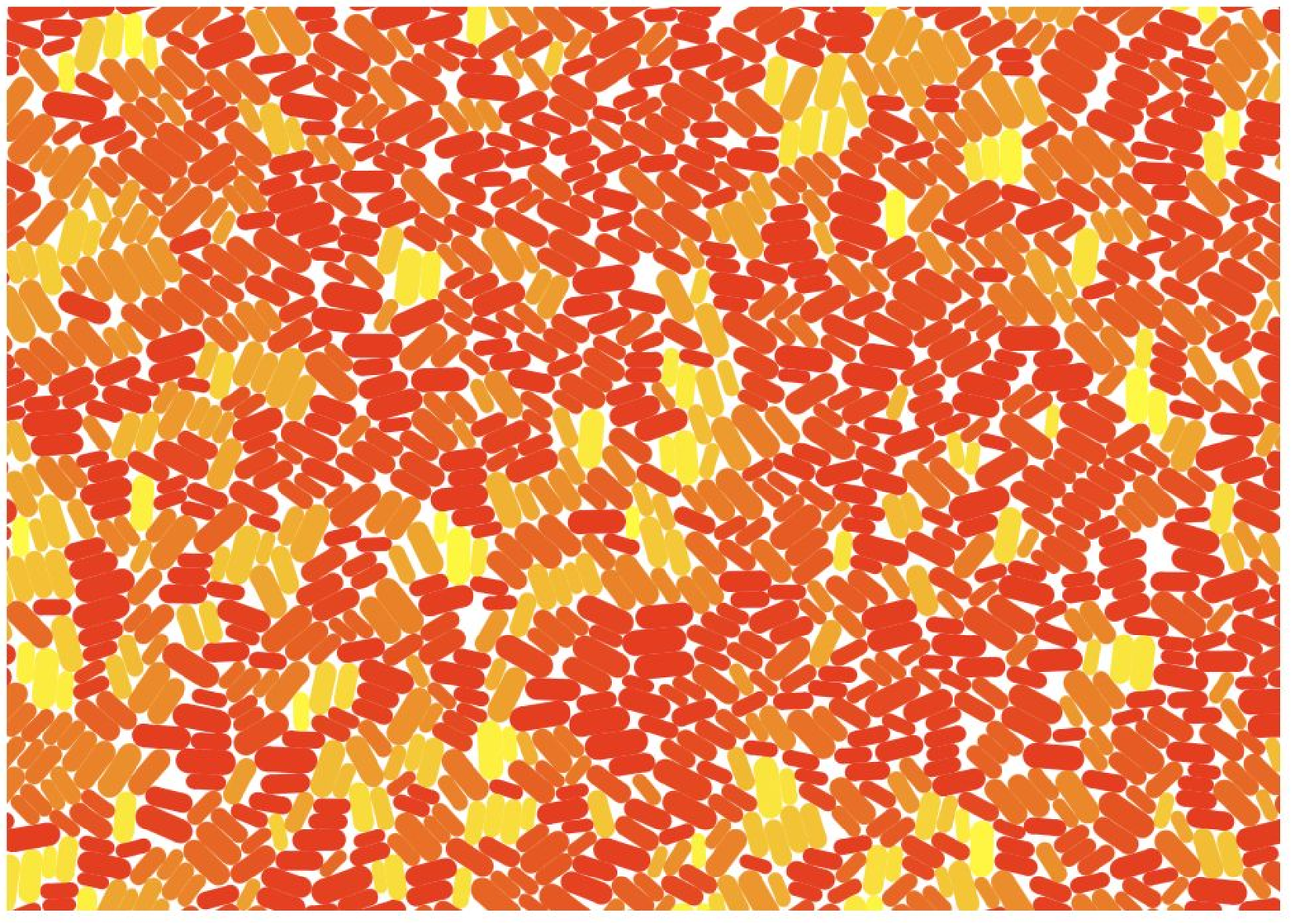}
\caption{(Color online) Color level map of particle orientations  
for $\eta=0.2$ (up) and $\eta=0.7$ (down) in the residual state.}
\label{sec:harmonic:carte_orientation_part}
\end{figure}

We plot the particle orientation anisotropy 
$a_p$ in Fig. \ref{sec:harmonic:orientation_part} as a function 
of $\eta$ at different stages of shear $\varepsilon_q$. We see that the particle 
orientations are isotropic for $\eta \leq 0.4$ at the initial state and they become 
increasingly anisotropic as $\eta$ increases beyond $0.4$. 
At nearly all stages of shear, 
$a_p$ is negative, and at most advanced stage, i.e. corresponding approximately 
to the residual state,  it becomes nearly independent of $\eta$. The large absolute value of 
$a_p$ ($\sim 0.35$) suggests that many particles are aligned in horizontal 
layers just as in nematic order. One example of this nematic ordering is 
shown in Fig. \ref{sec:harmonic:carte_orientation_part} for two different values of $\eta$. This ordering may be attributed to 
the favored mechanical equilibrium of the particles under the action of 
vertical stress and enhanced by boundary alignment of the elongated particles 
\cite{Nouguier-Lehon2003,Zuriguel2007,Zuriguel2008,Hidalgo2009}. This point will be 
analyzed more deeply below. 

\subsection{Particle connectivity}

The primary statistical quantity describing the contact 
network is the coordination number $z$ (average number of contacts per particle).
For our elongated particles, each side-to-side contact
is counted as one contact  even if  side-to-side contacts are treated 
as four point contacts belonging to 
the contact segment (see section \ref{procedure}). 
The floating particles with no force-bearing contact 
(i.e. with less than two active contacts) are thus removed from the statistics. 
The fraction of floating particles decreases linearly with $\eta$ 
from 17\% for $\eta=0$ to 10\% for $\eta=0.7$.

\begin{figure}
\includegraphics[width=8cm]{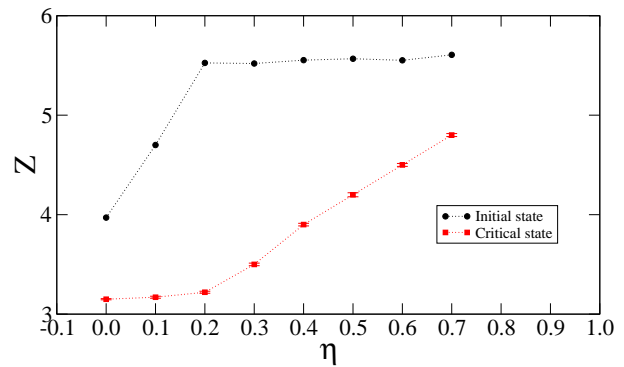}
\caption{Initial and residual coordination numbers  as a function of shape parameter $\eta$. The error bars represent the standard deviation 
in the residual state. 
\label{sec:harmonic:z_eta}}
\end{figure}

Figure \ref{sec:harmonic:z_eta} displays the evolution of $z$ as a 
function of $\eta$, in the initial and residual states. 
The initial-state value $z=4$ corresponds 
to a frictionless packing of circular particles in the 
isostatic state with $z=2d$ where $d$ is space 
dimension (indeed, the packings where prepared by setting the 
friction coefficient to zero). 
However, as $\eta$ increases, $z$ increases to a plateau value of 
$\sim 5.6$.  This is in agreement with recent work showing that 
large disordered jammed packings are isostatic only for disks or spheres 
\cite{Tkachenko1999,Roux2000a,Donev2004,
Donev2004a,Man2005,Donev2007,Yatsenko2007}.  
For nonspherical or noncircular particles, we have 
$z\leqslant d(d+1)$. 

Numerical results for   
frictional or frictionless systems of rigid or deformable disks 
and spheres \cite{Moukarzel1998,Edwards1998,Alexander1998,Tkachenko1999,Guises2009,Mailman2009}, 
as well as with ellipses and spheroids  \cite{Donev2004,Donev2004a,Man2005,Donev2007,Yatsenko2007}
confirm this point.
In the residual state, the mean value
of $z$ is below that in the isotropic state, and it  
grows  from $3$ to $5$ with $\eta$. It is interesting to 
note that $z$ does not follow the solid fraction which, as we have seen before,  is 
nonmonotonous as a function of $\eta$. This means that 
for large aspect ratios, the packings are loose but well 
connected. 
  
\begin{figure}
\includegraphics[width=8cm]{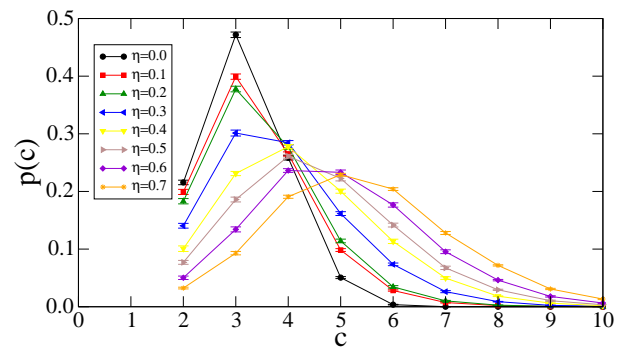}
\caption{Connectivity diagram for each samples expressing the 
fraction $P(c)$ of particles with exactly $c$ contacts in the residual state.
Note that the floating particles (i.e. with more than one active) are removed from the statistics
\label{sec:harmonic:P_c}}
\end{figure}

\begin{figure}
\includegraphics[width=9cm]{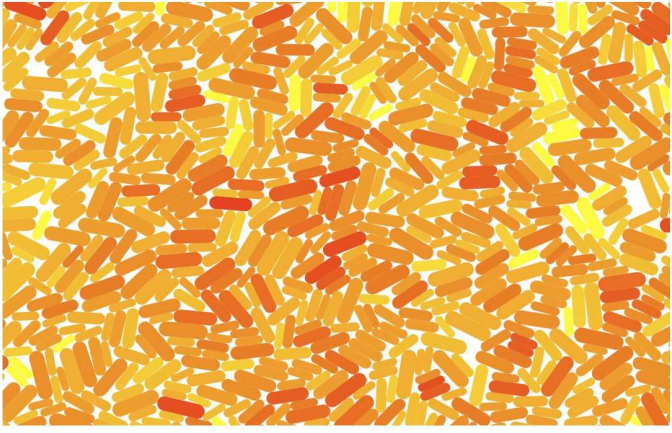}
\caption{(Color online) Color map  of particle connectivities. 
Color intensity is proportional to coordination number.  
\label{sec:Map_harmonic:P_c}}
\end{figure}

The connectivity of the contact network can be characterized in more detail by the 
fraction $P(c)$ of particles with exactly $c$ contact neighbors. 
Fig.  \ref{sec:harmonic:P_c} shows $P(c)$   in the residual state for 
different values of $\eta$. 
The distribution is increasingly broader as $\eta$ becomes larger.  
The particles can have as many as 10 contact neighbors 
at $\eta=0.7$. This is allowed both by the geometry and 
polydispersity of the particles as shown by a typical grey-level map of 
particle connectivities in Fig. \ref{sec:Map_harmonic:P_c}.   
For $\eta=0$,  we observe a peak at $c=3$. 
This peak slides gradually to $c=5$ at $\eta=0.7$ as observed also 
for $z$, which is, by definition, the mean value 
of $c$ for force-bearing contacts: $z=\langle c \rangle = \sum_{c=2}^\infty c\ P(c)$.

\subsection{Force and texture anisotropies}

Equation (\ref{eq:sigma}) shows that  
the expression of stress tensor  is an arithmetic mean involving the 
branch vectors $\bm \ell$ and contact force vectors $\bm f$.
This means that for the analysis of stress transmission and 
shear strength from a particle-scale viewpoint 
we need a statistical description of these quantities. 

\begin{figure}
\includegraphics[width=3.5cm]{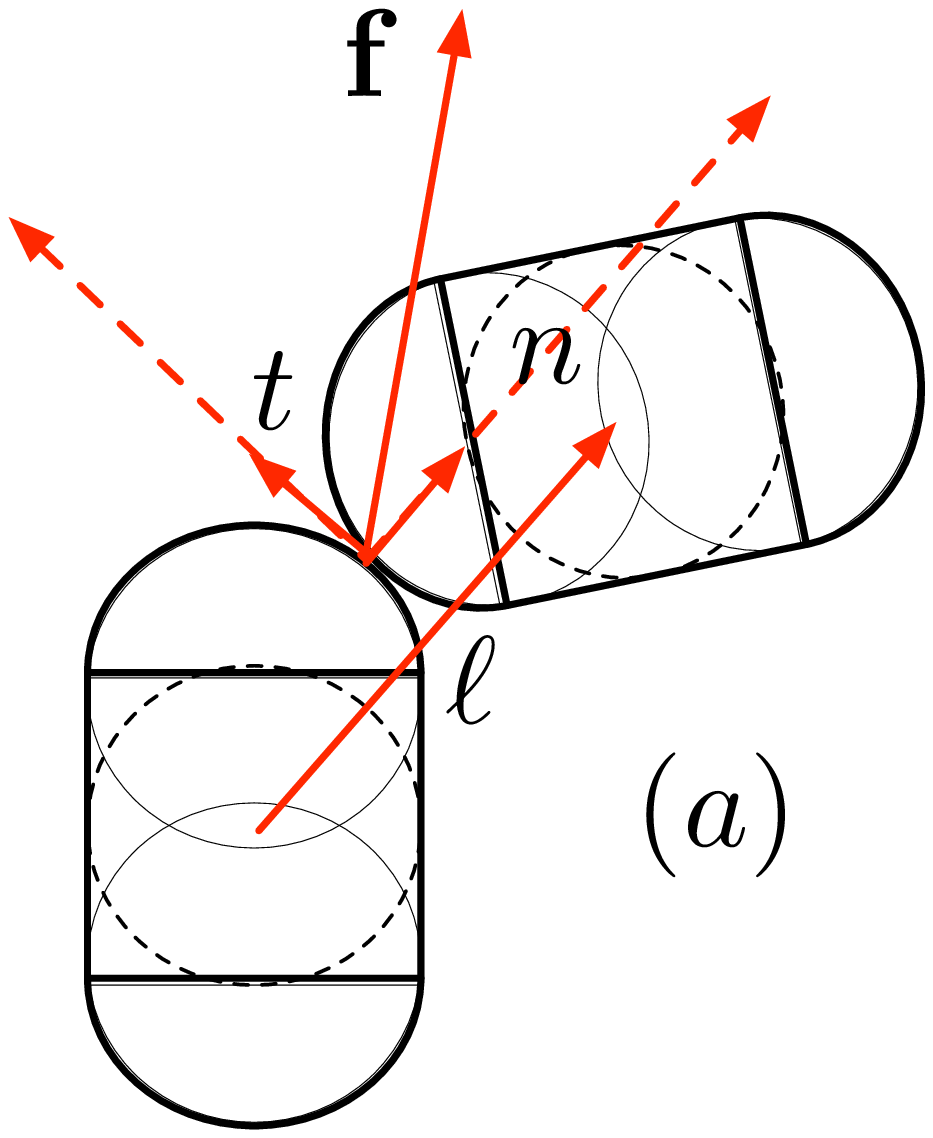}
\includegraphics[width=4.2cm]{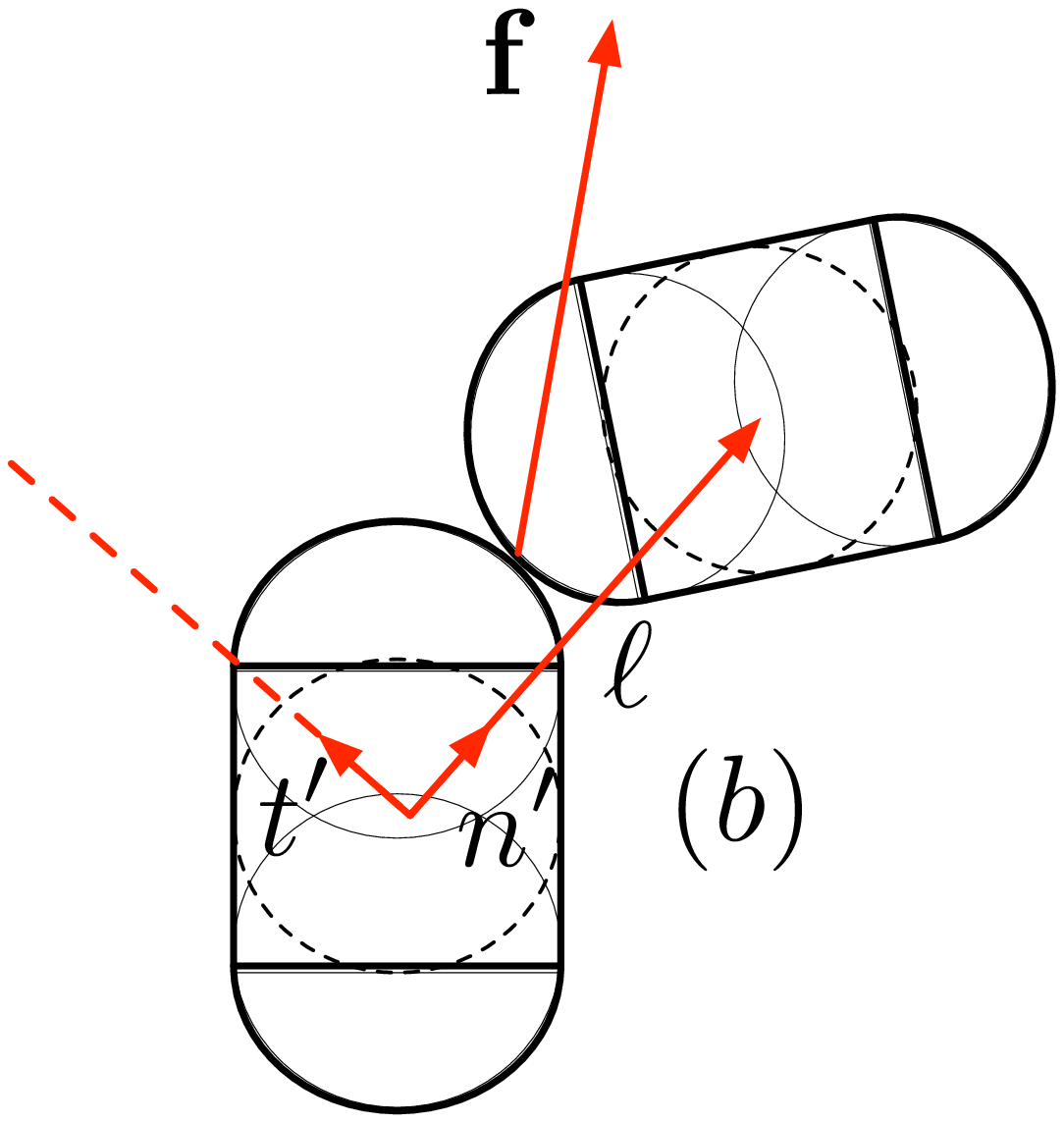}
\caption{Contact frame $(\bm n,\bm t)$ (a) and intercenter frame $(\bm{n'},\bm{t'})$ (b) \label{sec:numerical_procedure:frame}}
\end{figure}

A common approach used by various authors is to express branch vectors and 
contact force orientations
in terms of the contact direction, i.e. in the local {\it contact frame} $(\bm n, \bm t)$, 
where $\bm n$ is the unit vector perpendicular to the contact 
plane, and $\bm t$ is an orthonormal unit vector oriented along the tangential force
\cite{Satake1982,Bathurst1988,Rothenburg1989,Oda1999,Kruyt1996,Ouadfel2001,mirghasemi2002,Antony2004,
Kruyt2004,Cambou2004,Azema2007}
; see figure \ref{sec:numerical_procedure:frame}(a). The components  
of the branch vector and contact force are expressed in this frame as: 
\begin{equation}
\left\{
\begin{array}{lcl}
\bm \ell &=&  \ell_n \bm n + \ell_t \bm t,\\
\bm f &=& f_n \bm n + f_t \bm t,
\end{array}
\right.
\end{equation}
where $\ell_n$ and $\ell_t$ are the normal and tangential components of the branch vectors, and 
$f_n$ and $f_t$ the normal and tangential components of the contact force.

Remark that only for disks or spherical particles we have $\bm \ell = \ell \bm n$ 
where $\ell$ is the length of the branch vector.
A consequence of noncircular or nonspherical particle shape is to 
dissociate the contact frame from the {\it branch vector frame} $(\bm{n'}, \bm{t'})$, 
where $\bm{n'}$ is the unit vector along the branch $\bm \ell$  
and $\bm{t'}$ is the orthoradial unit vector  
\cite{Pena2007,Azema2009} ; see figure \ref{sec:numerical_procedure:frame}(b). 
We express the components  
of the branch vector and contact force also in this frame:
\begin{equation}
\left\{
\begin{array}{lcl}
\bm \ell &=&  \ell_{n'} \bm {n'} ,\\
\bm f &=& f_{n'} \bm {n'} + f_{t'} \bm {t'},
\end{array}
\right.
\end{equation}
where $f_{n'}$ and $f_{t'}$ are the {\it radial} and {\it orthoradial} components of the contact force, and 
$\ell_{n'}=\ell$.

\begin{figure}
\includegraphics[width=8cm]{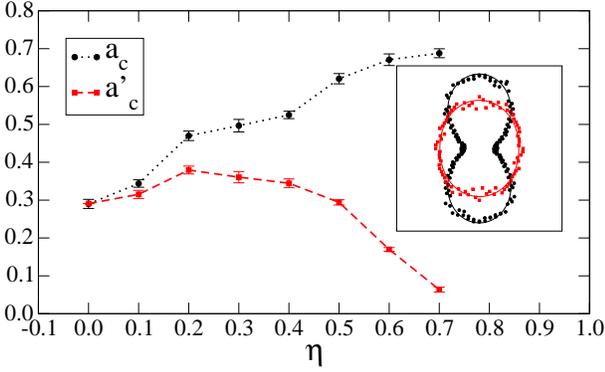}
\caption{Contact anisotropy $a_c$ and branch vector anisotropy $a'_c$ 
as a function of shape parameter $\eta$ averaged in residual state. The error bars represent the standard deviation 
in the residual state. The inset shows the angular probability densities $P_\theta(\theta)$ in black
and $P_{\theta'}(\theta')$ in red for $\eta=0.7$ calculated from the simulation 
data (points) together with the harmonic approximation (lines).
\label{sec:harmonic:a}}
\end{figure}

\begin{figure}
\includegraphics[width=8cm]{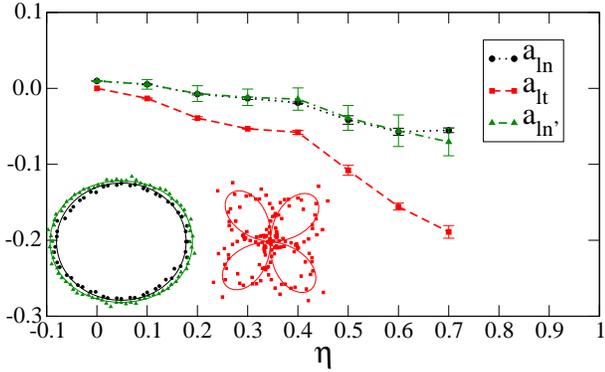}
\caption{Normal and tangential branch length anisotropies $a_{ln}$ and $a_{lt}$ and branch 
length anisotropy $a_{ln'}$ as a function of shape parameter $\eta$ in the residual state. 
The error bars represent the standard deviation 
in the residual state.
The inset  shows the angular average functions $\langle \ell_{ln} \rangle (\theta)$,  
$\langle \ell_{lt} \rangle (\theta) $  and $\langle \ell_{ln'} \rangle (\theta)$ in
black, red and green, respectively, for $\eta=0.7$ calculated from the 
simulation data (points) and approximated by harmonic fits  (lines).
\label{sec:harmonic:a_l}}
\end{figure}

\begin{figure}
\includegraphics[width=8cm]{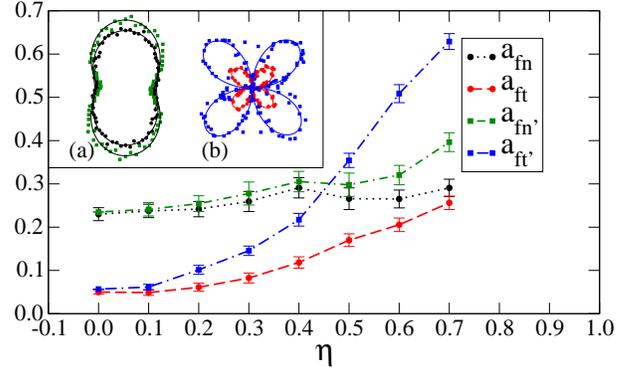}
\caption{Normal and tangential force anisotropies $a_{n}$ and $a_{t}$ 
and radial and orthoradial force anisotropies 
 $a'_n$ and $a'_t$ as a function of $\eta$ in the residual state.
 The error bars represent the standard deviation 
in the residual state.
The inset  shows the angular average functions $\langle f_{n} \rangle (\theta)$ 
 and $\langle f_{n'} \rangle (\theta) $ in black and green, respectively, (a) and 
$\langle f_{t} \rangle (\theta)$ and $\langle f_{t'} \rangle (\theta)$ in red and blue, respectively, (b)  
for $\eta=0.7$ calculated from the simulation data (points) together with the
harmonic approximation (lines). The error bars represent the standard deviation 
in the residual state.  
\label{sec:harmonic:a_n}}
\end{figure}

In two dimensions, let $\theta$ and $\theta'$ be the orientations of 
of $\bm n$ and $\bm{n'}$, respectively. From the numerical data, we can evaluate 
the probability density functions $P_\theta(\theta)$ and $P_{\theta'}(\theta')$ of 
contact and branch vector orientations, respectively, as well as the 
angular averages of the force components 
$\langle f_{n} \rangle (\theta)$, $\langle f_{t} \rangle (\theta)$, $\langle f_{n'} \rangle (\theta')$, 
and $\langle f_{t'} \rangle (\theta')$ and branch vector components  
$\langle \ell_{n} \rangle (\theta)$, $\langle \ell_{t} \rangle (\theta)$, 
$\langle \ell_{n'} \rangle (\theta')$.  
In the absence of an intrinsic polarity for  $\bm n$ and $\bm{n'}$, all these functions 
are $\pi$-periodic. 
The insets to Figs. \ref{sec:harmonic:a}, \ref{sec:harmonic:a_l} and \ref{sec:harmonic:a_n}
display polar representations of these functions for $\eta=0.7$ at the end of shearing. 
All these angular functions are generically anisotropic.
The peak values occur along the axis of compression ($\theta=\pi /2$) 
for $P_{\theta}$, $P_{\theta'}$, $\langle f_{n} \rangle$ and $\langle f_{n'} \rangle$, and along the 
axis of extension ($\theta=0$) for $\langle \ell_{n}\rangle$ and $\langle \ell_{n'}\rangle$.  
The maxima for the tangential components occur in the direction of  $\pi/4$ 
with respect to the axis of compression.  

The simple shapes of the above functions suggest that they can be 
approximated by their Fourier expansions up to the  
second term \cite{Rothenburg1989,Ouadfel2001,mirghasemi2002,Radjai2004a,Azema2009}:
\begin{equation}
\label{P_theta}
\left\{
\begin{array}{lcl}
P_\Theta (\Theta) &=& \frac{1}{2\pi}  \{ 1 + a^*_c \cos 2(\Theta - \Theta^*_c) \},  \\
\langle \ell_{n^*} \rangle (\Theta) &=&  \langle \ell \rangle    \{ 1 + a_{ln^*}\cos 2(\Theta - \Theta_{ln^*}) \} , \\
\langle \ell_{t^*} \rangle (\Theta) &=&  \langle \ell \rangle  a_{lt^*} \sin 2(\Theta - \Theta_{lt^*}) , \\
\langle f_{n^*} \rangle (\Theta) &=& \langle f \rangle  \{ 1 + a_{fn^*} \cos 2(\Theta - \Theta_{fn^*}) \}  \\
\langle f_{t^*} \rangle (\Theta) &=& \langle f \rangle  a_{ft^*} \sin 2(\Theta - \Theta_{ft^*}) , 
\end{array}
\right.
\end{equation}
where $\Theta$ stands either for $\theta$ or for $\theta'$ depending on the local frame 
used. The $\langle \ell \rangle$ is mean length of branch vectors and $\langle f \rangle$ is 
the mean force. 
$(a^*_c, a_{ln^*}, a_{lt^*}, a_{fn^*},a_{ft^*}) = (a_c, a_{ln}, a_{lt}, a_{fn},a_{ft})$ and 
$(\Theta^*_c,\Theta_{ln^*}, \Theta_{lt^*}, \Theta_{fn^*}, \Theta_{ft^*})=(\theta_c,\theta_{ln}, \theta_{lt}, \theta_{fn}, \theta_{ft})$ are the anisotropy parameters 
and the angle of privileged direction of each function in
the frame $(\bm n, \bm t)$. In the same way, we have $(a^*_c, a_{ln^*}, a_{lt^*}, a_{fn^*},a_{ft^*}) = (a'_c, a_{ln'}, a_{lt'}, a_{fn'},a_{ft'})$ and 
$(\Theta^*_c,\Theta_{ln^*}, \Theta_{lt^*}, \Theta_{fn^*}, \Theta_{ft^*})=(\theta'_c,\theta_{ln'}, \theta_{lt'}, \theta_{fn'}, \theta_{ft'})$ 
in the frame $(\bm{n'},\bm{t'})$. 

Note that, by construction, we have 
$a_{lt'}=0$ and $\theta_{lt'}=0$.
In the following, we will refer to $a_c$ as contact anisotropy, to $a'_c$ as 
branch vector orientation anisotropy,   
to $(a_{ln^*}, a_{lt^*})$ as branch length anisotropies 
and to $(a_{fn^*}, a_{ft^*})$ as normal and tangential or radial and orthoradial force 
anisotropies depending on the local frame \cite{Azema2007,Voivret2009}. These harmonic approximations 
are well fit to our data as shown in the insets to 
Figs. \ref{sec:harmonic:a}, \ref{sec:harmonic:a_l} and \ref{sec:harmonic:a_n}.

In practice, it is convenient to estimate the above anisotropies 
from the following {\it fabric and force tensors} \cite{Radjai1998}: 
\begin{equation}
\label{Chi_tensors}
\left\{
\begin{array}{lcl}
F^*_{\alpha \beta} &=&
\frac{1}{\pi} \int\limits_{0}^\pi   n^*_\alpha  n^*_\beta P_\Theta(\Theta)d \Theta ,  \\ 
\chi^{ln^*}_{\alpha \beta} &=& 
\frac{1}{\langle \ell \rangle} \int\limits_{0}^\pi  \langle \ell_{n^*} \rangle(\Theta)  n^*_\alpha  n^*_\beta P_ \Theta(\Theta) d \Theta,  \\ 
\chi^{lt^*}_{\alpha \beta} &=& 
\frac{1}{\langle \ell \rangle} \int\limits_{0}^\pi  \langle \ell_{t^*} \rangle(\Theta)  n^*_\alpha  t^*_\beta P_ \Theta(\Theta) d \Theta,  \\ 
\chi^{fn^*}_{\alpha \beta} &=& 
\frac{1}{\langle f \rangle} \int\limits_{0}^\pi  \langle f_{n^*} \rangle(\Theta)  n^*_\alpha  n^*_\beta P_ \Theta(\Theta) d \Theta,  \\ 
\chi^{ft^*}_{\alpha \beta} &=& 
\frac{1}{\langle f \rangle} \int\limits_{0}^\pi  \langle f_{t^*} \rangle(\Theta)  n^*_\alpha  t^*_\beta P_ \Theta(\Theta) d \Theta,  \\ 
\end{array}
\right.
\end{equation}      
where $\alpha$ and  $\beta$ design the components in the considered frame. 
Note that, by construction, 
we have  $\bm \chi^{lt'}_{\alpha \beta}=0$. From equations (\ref{P_theta}) and (\ref{Chi_tensors}), 
assuming that in a sheared state $\Theta^*_c=\Theta_{fn^*}=\Theta_{ft^*}=\Theta_\sigma$, 
$\Theta_{ln^*}=\Theta_{lt^*}=0$ or $\theta_\sigma$, the following 
relations are easily obtained:
\begin{equation}
\label{Aniso_values}
\left\{
\begin{array}{lcl}
a^*_c &=& 2(F^*_{1}-F^*_{2}) /(F^*_{1}+F^*_{2}),  \\ 
a_{ln^*} &=& 2(\chi^{ln^*}_{1} - \chi^{ln^*}_{2})/(\chi^{ln^*}_{1} + \chi^{ln^*}_{2}) - a^*_c,  \\ 
a_{lt^*}&=& 2(\chi^{l^*}_{1} - \chi^{l^*}_{2})/(\chi^{l^*}_{1} + \chi^{l^*}_{2}) - a^*_c - a_{ln^*},  \\ 
a_{fn^*}&=& 2(\chi^{fn^*}_{1} - \chi^{fn^*}_{2})/(\chi^{fn^*}_{1} + \chi^{fn^*}_{2}) - a^*_c,  \\
a_{ft^*}&=& 2(\chi^{f^*}_{1} - \chi^{f^*}_{2})/(\chi^{f^*}_{1} + \chi^{f^*}_{2}) - a^*_c - a_{fn^*},
\end{array}
\right.
\end{equation}      
where  $\bm \chi^{l^*} = \bm \chi^{ln^*} + \bm \chi^{lt^*}$, $\bm \chi^{f^*} = 
\bm \chi^{fn^*} + \bm \chi^{ft^*}$ and
the indices $1$ and $2$ refer to the principal values of each tensor. 
By construction, we have 
$(F^*_{1}+F^*_{2})=1$, $(\chi^{l^*}_{1} + \chi^{l^*}_{2})=\langle \ell \rangle$ and $(\chi^{f^*}_{1} + \chi^{f^*}_{2}) = 
\langle f \rangle$. Note that $a^*_c$, $a_{fn^*}$ and $a_{ft^*}$ are always positive whereas
$a_{ln^*}$ and $a_{lt^*}$ are negative. We have  
$\Theta_{ln^*}=0$ and $\Theta_{lt^*}=0$,.

Figure \ref{sec:harmonic:a} displays the variation of  
contact anisotropy  
$a_c$ and branch vector orientation anisotropy $a_c'$, 
both averaged in the residual state, as a function of $\eta$.
We observe  two distinct behaviors: 
$a_c$ increases quickly from $0.3$ to $0.7$ with $\eta$ whereas, after a slight increase,
$a_c'$ declines to nearly $0$ for $\eta=0.7$. 
It is often admitted that 
the shear strength in granular media is a consequence of 
the buildup of an anisotropic geometrical structure due to friction and steric exclusions 
between particles \cite{Oda1980,Satake1982,Cambou1993,Antony2004,Kruyt2004,Radjai2004a}.
But, here we have two different structural anisotropies $a_c$ and $a_c'$ that vary 
in opposite directions as $\eta$ is increases. Hence, when the granular structure is complex as 
in our packings of nonspherical particle shapes,  
the choice of the statistical representation of the granular structure has to be specified 
\cite{Azema2007,Azema2009}. 
This point will be addressed in more detail in section \ref{contact_type}.

The branch vector length anisotropies $a_{ln}$, $a_{lt}$ and $a_{ln'}$, 
averaged in the residual state,  are plotted in Fig.\ref{sec:harmonic:a_l}
 as a function of $\eta$. These parameters are negligibly small 
 at small values of $\eta$, i.e. for nearly circular particles, and 
 decline to negative values as $\eta$ is increased.  
 This means that the particles tend to form longer 
 branch vectors with their neighbors in the direction 
 of extension, suggesting that the particles touch 
 preferentially along their minor axes when the contact orientation is close to 
 the compression axis, and along their major axis   when the contact orientation is close to 
 the extension axis; see section \ref{contact_type}. 
It is also remarkable that $a_{ln}\simeq a_{ln'}$ whereas $|a_{ln}|<|a_{lt}|$ 
particularly for $\eta \geqslant 0.4$. 

The normal and tangential force 
anisotropies $a_{fn}$ and $a_{ft}$ are plotted in Fig. \ref{sec:harmonic:a_n} as 
a function of $\eta$, together with 
the radial and orthoradial force anisotropies $a_{fn'}$ and $a_{ft'}$, 
averaged in the residual state.
In contrast to contact anisotropy, we see that 
$a_{fn}$ and $a_{fn'}$ grow together slowly until $\eta=0.4$, 
then $a_{fn}$ remains nearly constant whereas $a_{fn'}$ increases.
On the other hand, the anisotropy $a_{ft'}$ of orthoradial forces  
grows much faster  with $\eta$ than the anisotropy $a_{ft}$ of tangential forces.
Remarkably, from $\eta>0.4$ the orthoradial force anisotropy is higher than 
radial force anisotropy ($a_{ft'}> a_{fn'}$) whereas, 
even if the tangential force anisotropy increases with $\eta$, 
it is still below the normal force anisotropy ($a_{ft}< a_{fn}$) 
and remains always below the contact anisotropy ($a_{fn}<a_c$). 
In other words, the force anisotropy described  in terms of 
branch vectors reflects more sensitively the effect of 
particle shape elongation than in terms of contact normal vectors. 
We will see below that this behavior is related 
to the mobilization of friction (section \ref{Geom_mec_origins}) and contact 
types (section \ref{contact_type}).

\section{Geometrical and mechanical origins of shear strength}
\label{Geom_mec_origins}

The stress tensor as formulated in Eq.(\ref{eq:sigma}) is a function of 
discrete microscopic parameters
attached to the contact network. For sufficiently large systems,  
the dependence of volume averages on individual discrete
parameters vanishes \cite{Landau1959,Rothenburg1989,Ouadfel2001} and 
the discrete sums can be replaced by integrals. 
According to Eq. (\ref{eq:sigma}), we have 
\begin{equation}
V \sigma_{\alpha\beta} =  \sum_{c \in i} f_{\alpha}^c r_{\beta}^c = N_c \langle f_\alpha \ell_\beta \rangle, 
\end{equation}
where $N_c$ is the total number of contacts.
By writing  the average on the right hand side in integral form, we get 
\begin{equation}
\sigma_{\alpha\beta} = n_c \int_\Omega   f_\alpha \ell_\beta \ P_{\ell f} d{\bm f} \ d{\bm \ell},
\label{eqn:isig}
\end{equation}
where $P_{\ell f}$ is the joint probability density of forces and branch vectors, 
$n_c$ is the number density of contacts and $\Omega$ is the integration domain 
in the space $(\bm \ell, \bm f)$.  

The integral appearing in equation (\ref{eqn:isig}) can be reduced by 
integrating first with respect to  the forces and branch vector lengths.  
Considering the components of the forces and branch vectors in one of the 
two local frames $(\bm n, \bm t)$ or $(\bm n', \bm t')$ and neglecting 
the branch vector-force correlations, we get \cite{Rothenburg1989,Azema2009,Radjai2009a,Voivret2009}:
\begin{eqnarray}
\sigma_{\alpha\beta} &=& n_c \int\limits_0^{\pi}  
\{ \langle \ell_{n^*} \rangle (\Theta) \ n^*_\alpha (\Theta) + \langle \ell_{t^*} \rangle (\Theta) \ t^*_\beta (\Theta) \} \nonumber \\
 & & 
\{ \langle f_{n^*} \rangle (\Theta) \ n^*_\alpha (\Theta) + \langle f_{t^*} \rangle (\Theta) \ t^*_\beta (\Theta) \}  P(\Theta) \ d\Theta. \nonumber \\
& &
\label{eqn:isig2}
\end{eqnarray}

The expression of the stress tensor by equation (\ref{eqn:isig2}) makes appear explicitly 
the average angular functions representing the fabric and force states. 
Using the harmonic approximation (\ref{P_theta}), equation (\ref{eqn:isig2}) can be 
integrated with respect to space direction $\Theta$ and the stress invariants 
$p$ and $q$ extracted. Assuming that the stress tensor is coaxial 
with the fabric and force tensors (\ref{Chi_tensors}), we get 
the following simple relations:      

\begin{equation}
\frac{q}{p} \simeq
\left\{
\begin{array}{ll}
\frac{1}{2} (a_c + a_{ln} + a_{lt} + a_{fn} + a_{ft} ) & \mbox{ in $(\bm n, \bm t)$} \\
\\
\frac{1}{2} (a'_c + a_{ln'}  + a_{fn'} + a_{ft'} ) & \mbox{ in $(\bm{n'}, \bm{t'})$}. \\
\end{array}
\right.
\label{q_p_aniso2}
\end{equation}      
The assumption of coaxiality is natural since, even if the preferential 
orientations of the forces and branch vectors are not fully correlated,  
we observe that shearing tends to align the contacts and forces with the 
principal directions of the stress tensor.     

\begin{figure}
\includegraphics[width=8cm]{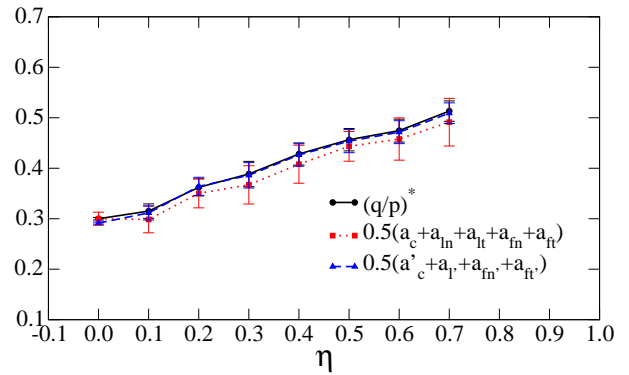}
\caption{Normalized shear stress $(q/p)^*$ in the residual state 
as a function of $\eta$ together with two analytical expressions given by 
equation (\ref{q_p_aniso2}). The error bars represent the standard deviation 
in the residual state.}
\label{sec:harmonic:q_p}
\end{figure}

Figure \ref{sec:harmonic:q_p} displays the residual-state value 
of the normalized shear stress $(q/p)^*$  as a function 
of $\eta$ calculated both directly from the simulation data and from 
equation (\ref{q_p_aniso2}) separately for the two local frames by using the 
values of various anisotropies estimated from the simulation data. 
As we see, for both local frames, equation  (\ref{q_p_aniso2}) 
provides an excellent fit to the data for all values of $\eta$. 
Note, however, that the second expression in equation  (\ref{q_p_aniso2})  
is more simple than the first expression 
(4 anisotropic parameters vs 5 anisotropic parameters) and the resulting fit 
appears to be more accurate.

The two equations (\ref{q_p_aniso2}) are interesting as they reveal 
distinct origins of shear strength in terms of force and texture 
anisotropies with two different decompositions. 
Various anisotropies do not contribute equally to 
the shear strength.  
The dominant anisotropies are texture anisotropies for projection 
on contact frame since $a_c+a_{ln}+a_{lt} > a'_c + a_{ln'}$ 
and force anisotropies for projection on the branch vector frame since 
$a_{fn'}+a_{ft'} > a_{fn} + a_{ft}$. The fact that the texture anisotropy prevails 
in the contact frame may be attributed to its strong correlation with particle 
orientations. Geometrically, for a particle oriented along a direction 
$\vartheta$, more contacts can be formed with the flat side of the particle 
with normals oriented along $\vartheta+\pi/2$ than with its rounded caps. This is 
consistent with the observation that the  
particle orientations are strongly anisotropic with an anisotropy $a_p$ of 
negative sign (preferred direction along 
the extension axis) and the contact normal anisotropy $a_c$ is positive (along 
the compression axis) and increases with aspect ratio; see section \ref{harmonics}. 

\begin{figure}
\includegraphics[width=9cm]{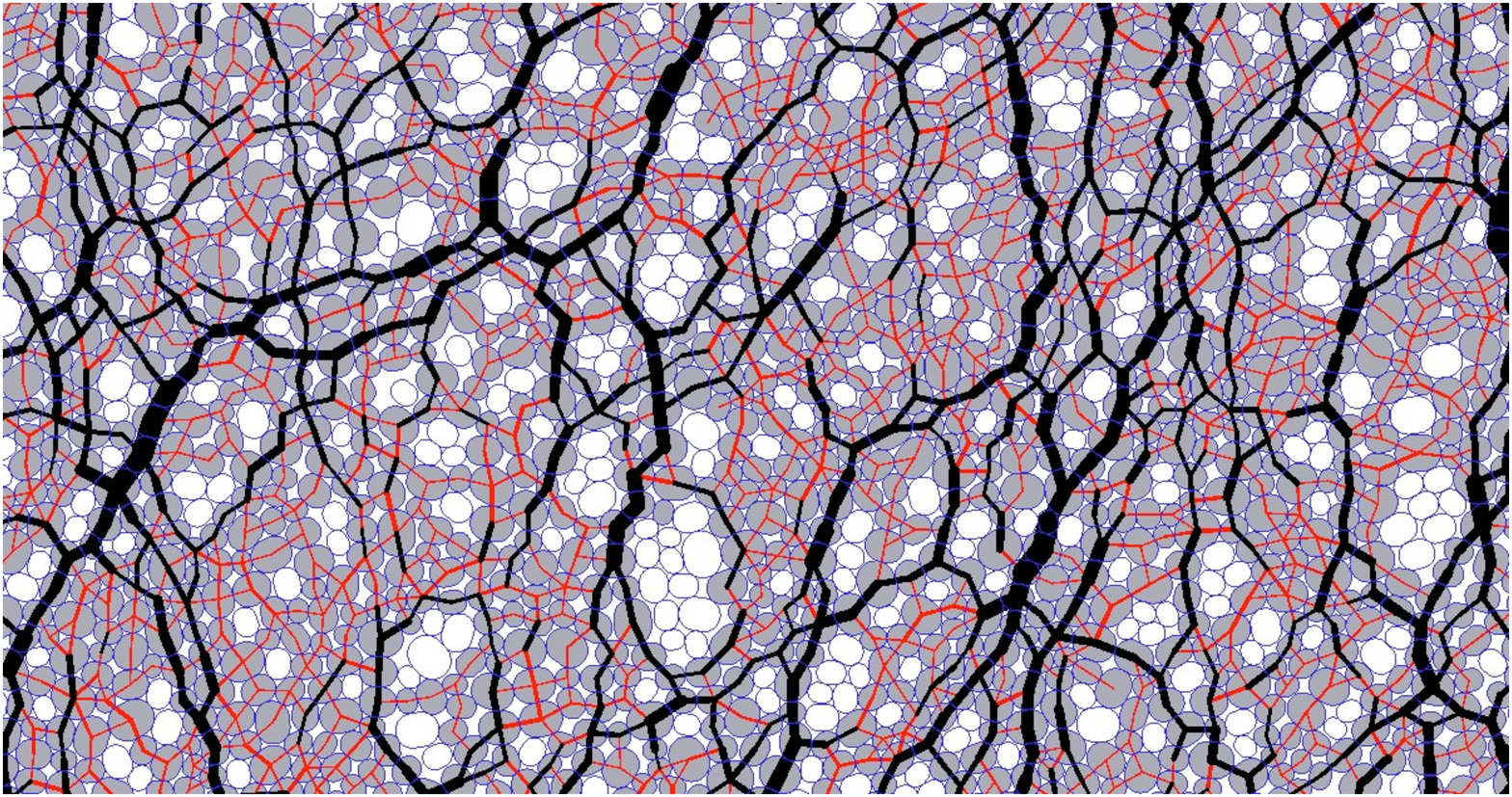}
\includegraphics[width=9cm]{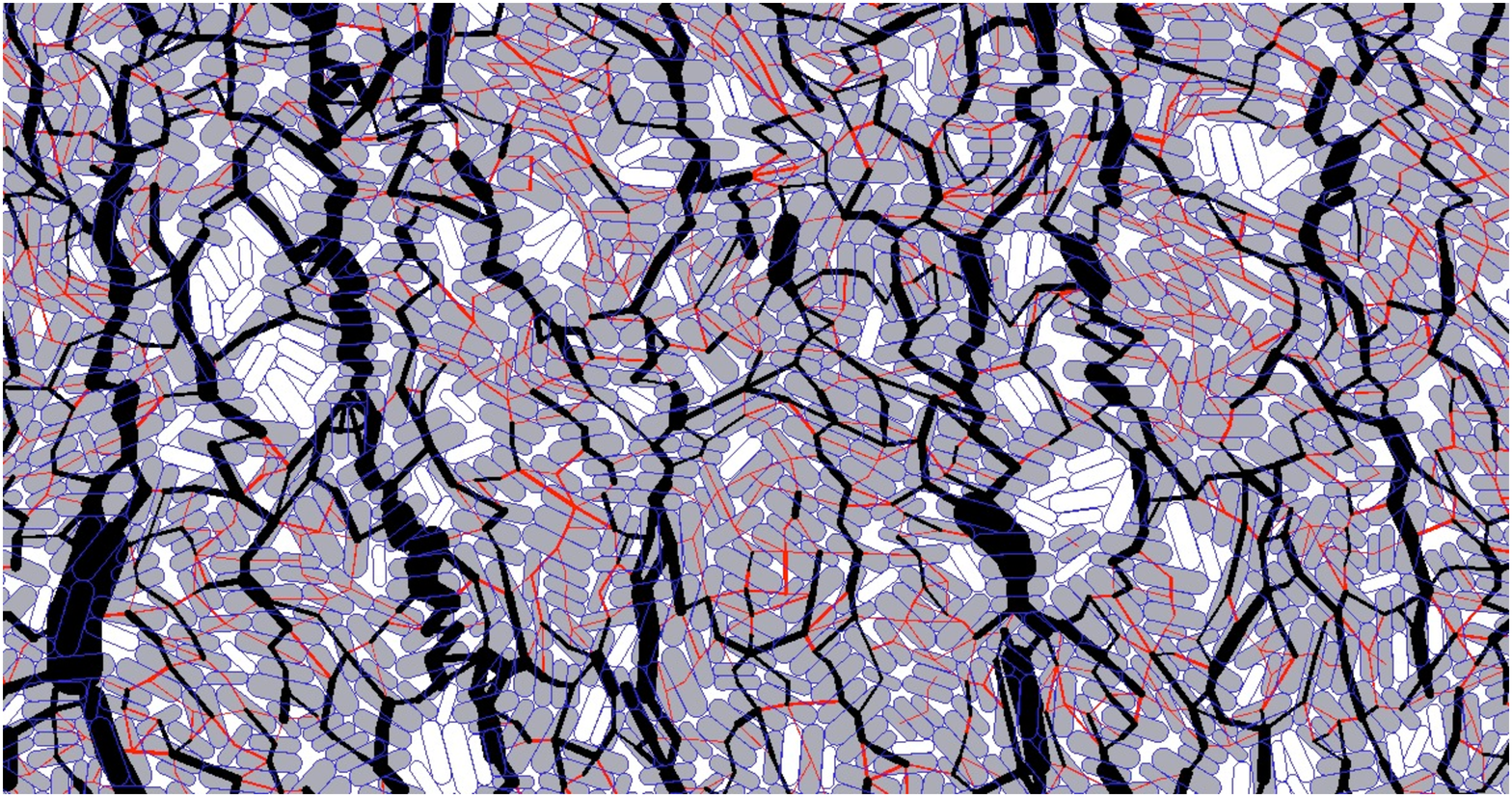}
\caption{Map of radial forces for $\eta=0.2$ (up) and $\eta=0.7$ (down). Line thickness is proportional to the radial force. We represent the strong network in black and the weak network in red lines (see section \ref{distribution}).
The floating particles excluded from the force network are in white.
\label{sec:harmonic:map_force}}
\end{figure}

\begin{figure}
\includegraphics[width=8cm]{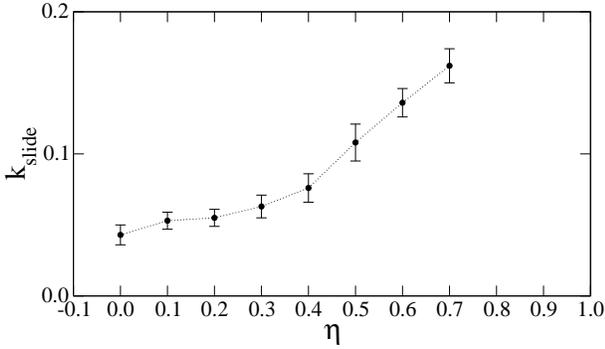}
\caption{Proportion of sliding contacts as a function of $\eta$ averaged in the residual state.
The error bars represent the standard deviation 
in the residual state.
\label{sec:harmonic:k_slide}}
\end{figure}

In Fig. \ref{sec:harmonic:map_force} two
maps of radial forces are shown  for packings with $\eta=0.2$ and $\eta=0.7$, respectively. 
In the presence of long parallel sides, the strong force 
chains are more tortuous. Hence, 
the stability of such structures requires strong activation 
of tangential forces. 
Indeed, we remark that the orthoradial force anisotropy is above the radial force  anisotropy             
 ($a_{ft'}>a_{fn'}$) for the most elongated particles in contrast to 
 the tangential force anisotropy which is below the normal force anisotropy 
 ($a_{ft}<a_{fn}$). As a result of the increasing activation  of tangential forces, 
 the fraction of sliding contacts (i.e. contacts where $|f_t|=\mu|f_n|$) grows with $\eta$ as shown 
 in Fig. \ref{sec:harmonic:k_slide}. The contributions of side-to-side and cap-to-side 
 contacts to force anisotropy and friction mobilization,  
 which are major effects of particle shape, will be analyzed 
 in section   \ref{contact_type}.

\section{Force distributions}
\label{distribution}
The force chains and spatially inhomogeneous stress distributions are 
well-known features of granular media. A well-known observation is
that a large number of contacts transmit very weak forces, a signature of 
the arching effect, whereas a smaller fraction of contacts carry strong 
force chains \cite{Radjai1998}. 
Force transmission  has been investigated by experiments and numerical  simulations for 
disks, ellipses and polygonal particles in 2D 
as for spherical, cylindrical and polyhedral particles in 3D\cite{Liu1995a,Mueth1998a,Radjai1996,Lovol1999,Bardenhagen2000,Antony2001,Silbert2002,Majmudar2005, Azema2007,Zuriguel2007,Metzger2008,Azema2009}. In close correlation with shear 
strength and solid fraction, 
the stress transmission is strongly influenced by particle shape. 
In particular, one expects that elongated particle shapes will influence mainly 
the distribution of weak forces by enhancing the arching effect.  

\begin{figure}
\includegraphics[width=8cm]{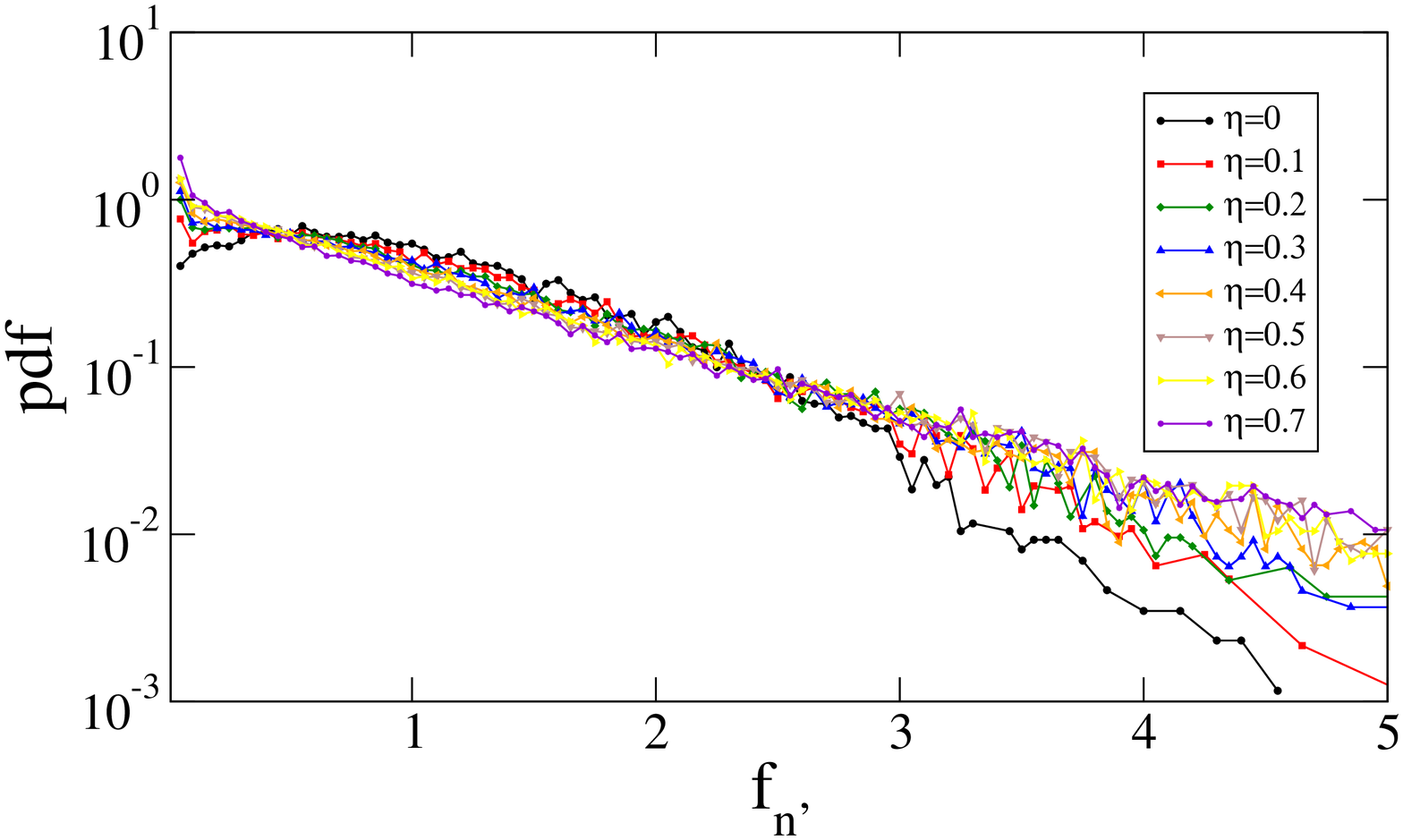}
\includegraphics[width=8cm]{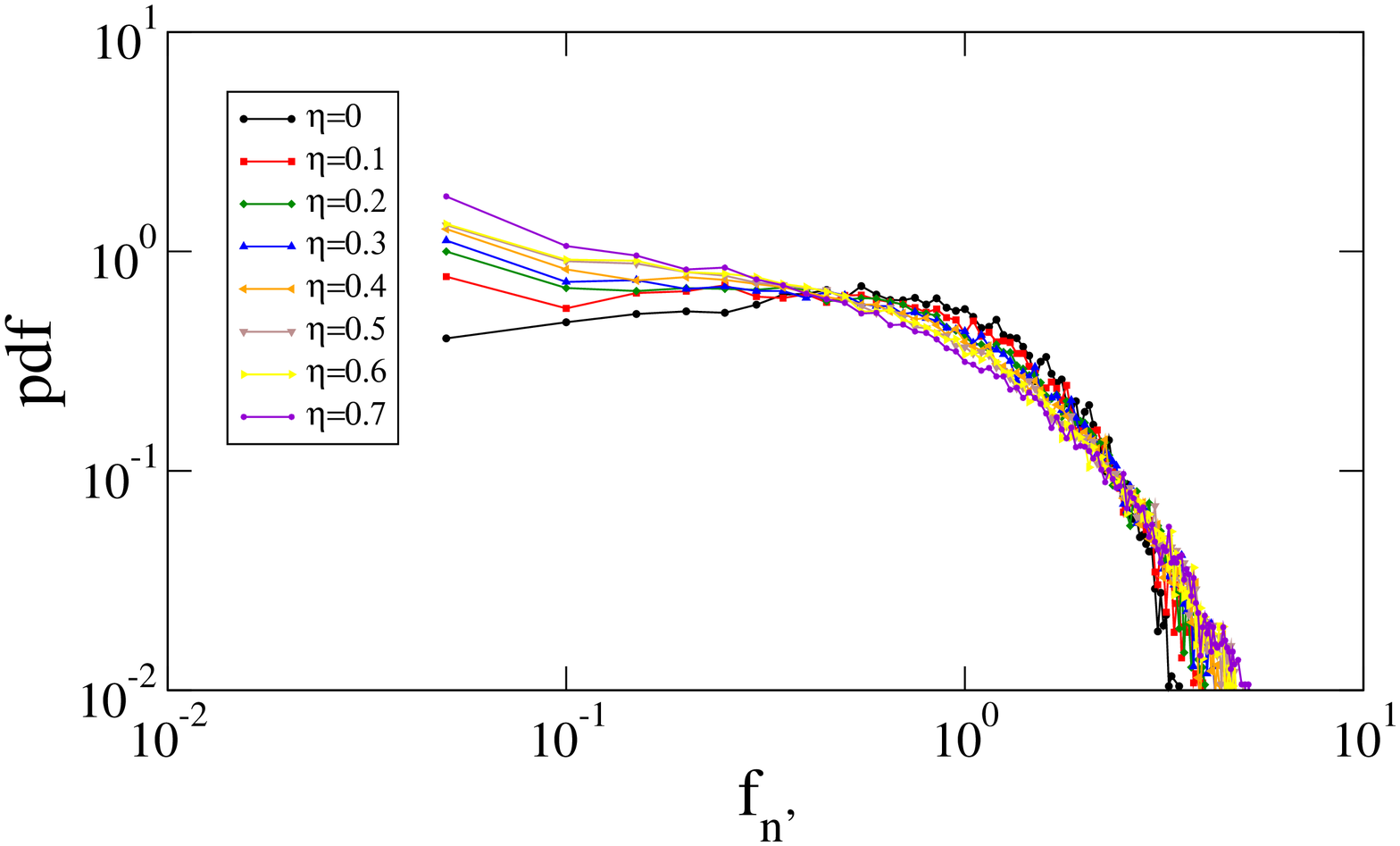}
\caption{Probability distribution function of radial forces $f_{n'}$
normalized by the average radial force $\langle f_{n'} \rangle$ in log-linear (up) 
and log-log (down) plots for different values of  $\eta$.
\label{sec:weakstrong:pcs_fn}}
\end{figure}

\begin{figure}
\includegraphics[width=8cm]{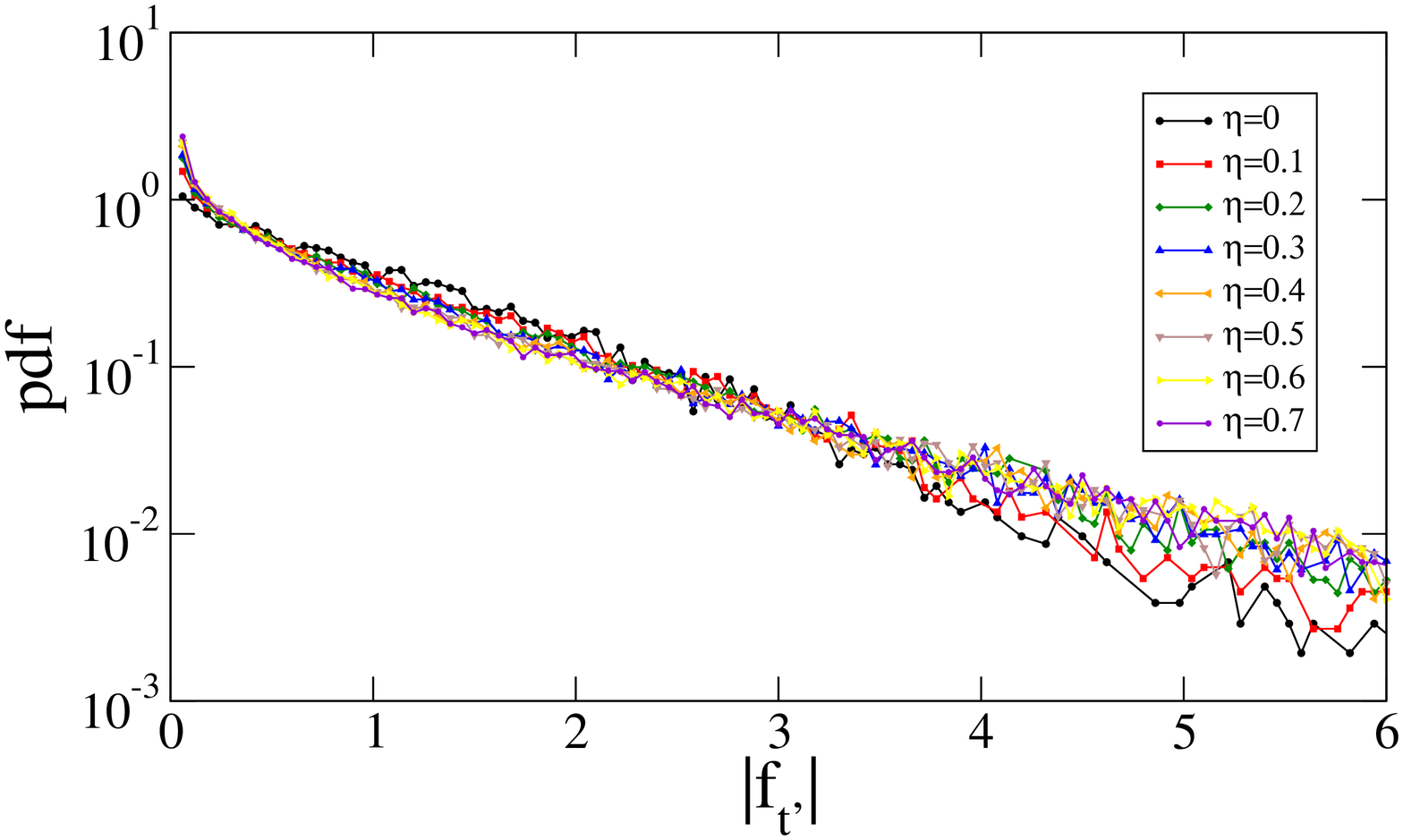}
\includegraphics[width=8cm]{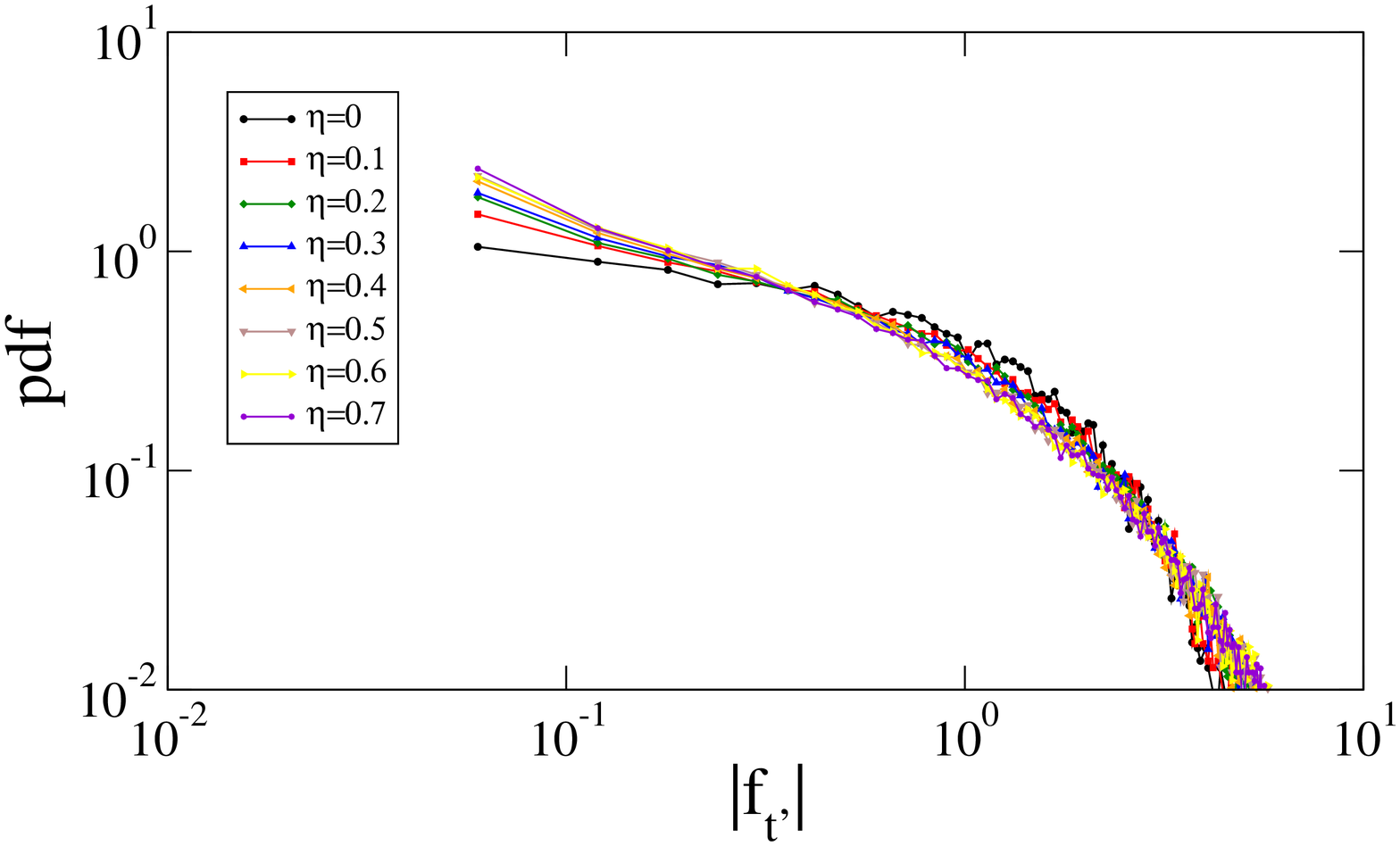}
\caption{Probability distribution function of orthoradial forces $f_{t'}$
normalized by the average orthoradial force $\langle |f_{t'}| \rangle$ 
in log-log for all values of $\eta$.
\label{sec:weakstrong:pcs_ftlog}}
\end{figure}

The probability density function (pdf) of radial forces normalized by the mean 
radial force $\langle f_{n'} \rangle$
is shown in Fig. \ref{sec:weakstrong:pcs_fn} in log-linear and log-log scales at 
large strains (the data are cumulated from several snapshots in the residual state) for 
all values of $\eta$.
As usually observed, in all packings the number of  forces 
above the mean $\langle f_{n'} \rangle$ falls off 
exponentially whereas the number of forces below the mean vary as a power-law:
\begin{equation}
P(f_{n'}) \propto
\left\{
\begin{array}{lcr}
e^{-\alpha_{n'}(\eta)  (1-f_{n'} / \langle f_{n'} \rangle}) &, & f_{n'}> \langle f_{n'} \rangle , \\
\Big(\frac{f_{n'}}{\langle f_{n'} \rangle}\Big)^{\beta_{n'}(\eta)} &, & f_{n'}< \langle f_{n'} \rangle,
\end{array}
\right.
\label{eqn:strong_fn}
\end{equation}      
where  $\alpha_{n'}(\eta)$ and $\beta_{n'}(\eta)$ are the exponents which 
decrease with $\eta$ from $\alpha^{n'}(0) \simeq 1.69$ to $\alpha^{n'}({0.7}) \simeq 0.88$, 
and from $\beta^{n'}(0) \simeq 0.13$ to $\beta^{n'}({0.7})\simeq -0.49$.   
Figure \ref{sec:weakstrong:pcs_ftlog} shows the pdf $P(f_{t'})$ 
of orthoradial forces normalized by by the mean orthoradial 
force $\langle f_{t'} \rangle$ in each packing. These distributions are
also characterized by an exponential falloff for the forces above the 
average force $\langle f_{t'} \rangle$ and a 
power law for the forces below $\langle f_{t'} \rangle$: 
\begin{equation}
P(f_{t'}) \propto
\left\{
\begin{array}{lcr}
e^{-\alpha_{t'}(\eta)  (1-|f_{t'}| / \langle |f_{t'}| \rangle}) &, & |f_{t'}|> \langle |f_{t'}| \rangle , \\
\Big(\frac{|f_{t'}|}{\langle |f_{t'}| \rangle}\Big)^{\beta_{t'}(\eta)} &, & |f_{t'}|< \langle |f_{t'}| \rangle,
\end{array}
\right.
\label{eqn:strong_ft}
\end{equation}      
with the corresponding exponents $\alpha_{t'}(\eta)$ and $\beta_{t'}(\eta)$, which decrease 
from $\alpha^{t'}(0) \simeq 1.09$ to $\alpha^{t'}({0.7}) \simeq 0.73$, 
and  from $\beta^{t'}(0) \simeq -0.37$ to $\beta^{t'}({0.7})\simeq -0.73$.

These distributions show clearly the larger inhomogeneity 
of stress transmission in a granular packing composed of elongated particles.
We find that (the results not shown here), as with circular particles, 
the contacts can be classified 
into strong and weak networks. 
Evaluating $q/p$ separately for each network, it is found that the shear 
stress is almost totally sustained by the strong contact 
network.

Maps of strong and weak radial networks for radial forces are displayed 
in Fig. \ref{sec:harmonic:map_force} 
for $\eta=0.2$ and $\eta=0.7$.  
The fraction of floating particles (less than two contacts) decreases 
linearly with $\eta$ from 17\% for $\eta=0$ to 10\% for $\eta=0.7$.
Hence, more particles are involved in the contact network for more elongated 
particles, but it is remarkable that   
 the proportion of weak forces grows from $60\%$ for $\eta=0$  
 to $70\%$ for $\eta=0.7$. In other words,
although the number of strong contacts decreases with $\eta$, stronger force chains 
occur with more elongated particles. Although we focused here on the networks of radial force components, we basically obtain the same conclusions for the normal force components. 

We also remark that the packings are increasingly
more inhomogenous in the sense that as particle elongation increases, the packing involves 
less strong force chains in number but with stronger forces. This decreasing force homogeneity 
in spite of increasing connectivity (fig. \ref{sec:harmonic:z_eta}), means that force distributions
are controlled by more subtle details of the microstructure than the density of contacts
or solid fraction.
As we shall see below, this is related to the role of various contact 
types in the contact network.

\section{Effect of contact types}
\label{contact_type}

Remembering that RCR particles are clumps of two disks with one 
rectangle, in this section we  revisit the results of the previous sections in the light 
of the organization of cap-to-cap ($cc$), cap-to-side
($cs$) and side-to-side ($ss$) contacts in each packing. 
The side-to-side or side-to-cap contacts do not transmit torques, 
but they are able to accommodate force lines that are usually unsustainable 
by cap-to-cap contacts. For this reason, it is worth while 
trying to isolate their respective roles with respect to the shear strength. 
The proportions of these contact types and their contributions to 
the structural anisotropy and force transmission are key quantities 
for understanding the 
effect of particle shape on the shear strength properties of granular media \cite{Azema2007,Azema2009}.  

\begin{figure}
\includegraphics[width=8cm]{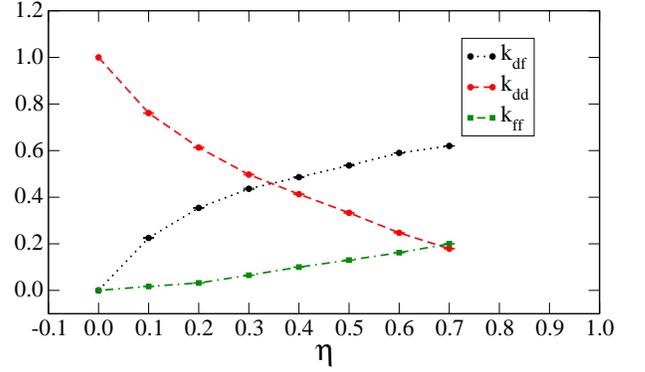}
\caption{Proportions of side-to-side ($ss$), cap-to-side ($cs$) and cap-to-cap ($cc$) 
contacts as a function of $\eta$ in the residual state. The error bars represent the standard deviation 
in the residual state.
\label{sec:TYPE_CONTACT:cc_ff_cs}}
\end{figure}

In the residual state, the proportions of different  contact types are nearly 
constant. Fig.  \ref{sec:TYPE_CONTACT:cc_ff_cs} shows the 
proportions $k_{cc}$, $k_{cs}$ and $k_{ss}$
of $cc$, $cs$ and $ss$ contacts  averaged
over the residual state as a function of $\eta$. We see that $k_{cc}$ declines 
with $\eta$ from 1 (for disks) to $0.2$ for $\eta=0.7$. At the 
same time, $k_{cs}$ and $k_{ss}$  increase  from $0$ to $0.6$ and to  0.2, respectively.  
Interestingly, $k_{cs}\simeq k_{cc}$ for 
$\eta \simeq 0.4$, and $k_{ss} \simeq k_{cs}$ for $\eta=0.7$. 
In this way, as the particle elongation increases, the packing passes from 
a contact network dominated by $cc$ contacts to a contact network 
dominated by the {\it complex} contacts $cs$ and $ss$.

\begin{figure}
\includegraphics[width=8cm]{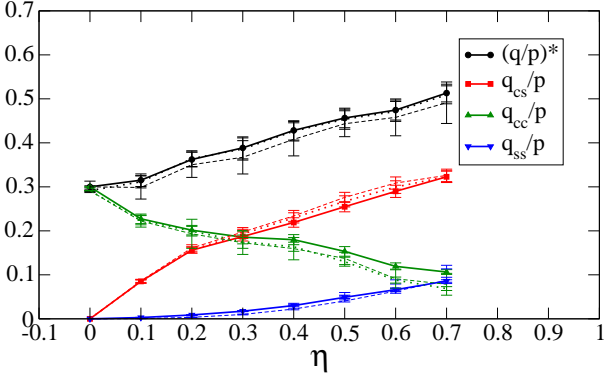}
\caption{Shear strength $(q/p)^*$ for $cs$, $ss$ and $cc$ contacts as a 
function of $\eta$, together with the harmonic approximation fits 
in $(\bm n,\bm t)$ frame
(-\;-\;-) and $(\bm {n'},\bm {t'})$ frame (...).
\label{sec:TYPE_CONTACT:ksi_qp}}
\end{figure}

To identify the impact of each contact type on the shear strength, 
we proceed by additive decomposition of the stress tensor by considering the 
expression (\ref{eqn:isig}) of the stress tensor and grouping the contacts according 
their types:
\begin{equation}
{\bm \sigma} = {\bm \sigma}_{cc} + {\bm \sigma}_{cs} + {\bm \sigma}_{ss},
\label{eqn:sigmacc_cs_ff}
\end{equation}
where  ${\bm \sigma}_{cc}$, ${\bm \sigma}_{cs}$ and ${\bm \sigma}_{ss}$ 
are obtained from the expression
of the stress tensor Eq. (\ref{eqn:isig}) by restricting the summation to $cc$, 
$cs$ and $ss$ contacts, respectively. The corresponding stress deviators  
$q_{cc}$, $q_{cs}$ and $q_{ss}$ are then calculated and normalized
by the mean pressure $p$. Fig. \ref{sec:TYPE_CONTACT:ksi_qp} shows 
$q_{cc}/p$,  $q_{cs}/p$ and $q_{ss}/p$ averaged in the residual state 
as a function of $\eta$. We see clearly that $q_{cc}/p$ follows 
a trend opposite to that of $q_{cs}/p$. For $\eta<0.3$, $(q/p)^*$ is dominated 
by $cc$ contacts. For
$\eta \simeq 0.3$, $cc$ and $cs$ contacts participate equally to the shear stress, 
and for $\eta>0.3$, the $cs$ contacts dominate $(q/p)^*$. 
Remarkably, $q_{ss}/p\simeq0$ for $\eta<0.4$.  
As we shall see below, the $ss$ contacts participate to the strong force 
chains only in the case of the most elongated particles. 
In this way, the growth of the number of $cs$ and $ss$ contacts 
shown in Fig. \ref{sec:TYPE_CONTACT:cc_ff_cs}
is clearly at the origin of a gradual consolidation of the packings 
as $\eta$ increases.

\begin{figure}
\includegraphics[width=8cm]{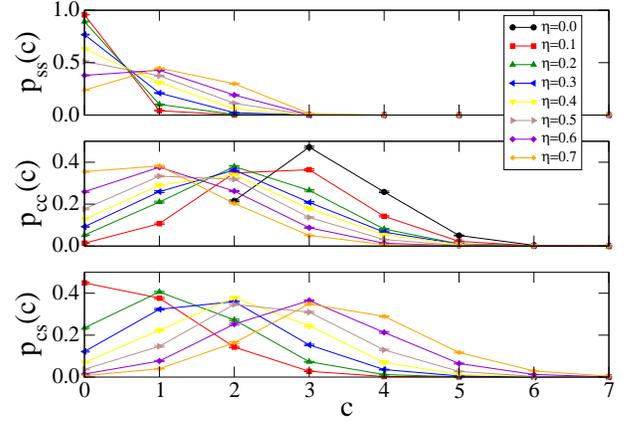}
\caption{Partial connectivity diagrams for all packings  in the residual state.
\label{sec:TYPE_CONTACT:P_c_cc}}
\end{figure}

In order to get further insight into the organization of different contact types, it 
is useful to consider partial connectivities $P_{cs}(c)$,  $P_{cc}(c)$ and $P_{ss}(c)$ 
defined as the fraction of particles with exactly $c$ contacts of $cs$ type, $cc$ type and 
$ss$ type.  These functions are displayed in Fig. \ref{sec:TYPE_CONTACT:P_c_cc} 
for all our packings. 
Note that, by definition $P_{cc}(c) \equiv P(c)$ for $\eta=0$.
We see that $P_{cc}$ gets narrower as $\eta$ increases whereas $P_{cs}$ and $P_{ss}$
get broader. It should be noted that, even for the most elongated particles, 
a particle can have at most two $ss$. This means that, the elongated particles
tend mainly to pile up like bricks. On the other hand, the peak of  $P_{cs}$ 
slides to the larger values as $\eta$ increases. For $\eta=0.7$, most particles 
have three or four $cs$ contacts (for nearly $40\%$).  

We now  consider the anisotropy of the branch vectors and contact forces 
supported by the three contact types at the contact and branch frames.
Following the same procedure as for  the stress 
tensor (see equation (\ref{eqn:sigmacc_cs_ff}), we perform an additive decomposition of  
the fabric and force tensors:
\begin{equation}
\label{Chi_tensors_cs_cc_ff}
\left\{
\begin{array}{lcl}
\bm F^* &=& \bm F^*_{cc} + \bm F^*_{cs} +  \bm F^*_{ss},  \\ 
\bm \chi^{ln^*}&=& \bm \chi^{ln^*}_{cc} + \bm \chi^{ln^*}_{cs} + \bm \chi^{ln^*}_{ss},  \\ 
\bm \chi^{fn^*}&=& \bm \chi^{fn^*}_{cc} + \bm \chi^{fn^*}_{cs} + \bm \chi^{fn^*}_{ss},  \\ 
\bm \chi^{ft^*}&=& \bm \chi^{ft^*}_{cc} + \bm \chi^{ft^*}_{cs} + \bm \chi^{ft^*}_{ss},  \\ 
\end{array}
\right.
\end{equation}      
where the indices refer to the partial contributions of $cc$, $cs$ and $ss$ contacts.
The corresponding anisotropies of each tensor can be extracted. 
In principle, the principal directions of these partial
tensors do not coincide with those  of the overall tensors at all stages of shearing. 
But, in practice, in the residual state, the principal directions coincide so that 
the global anisotropy of each tensor is the sum of its partial 
anisotropies: 
\begin{equation}
\frac{q_\gamma}{p} \simeq
\left\{
\begin{array}{ll}
\frac{1}{2} (a_{c\gamma} + a_{ln\gamma} + a_{lt\gamma} + a_{fn\gamma} + a_{ft\gamma} ) & \mbox{ in $(\bm n, \bm t)$ } \\
\\
\frac{1}{2} (a'_{c\gamma} + a_{ln'\gamma}  + a_{fn'\gamma} + a_{ft'\gamma} ) & \mbox{ in $(\bm{n'}, \bm{t'})$, } \\
\end{array}
\right.
\label{q_p_aniso3}
\end{equation}      
where $\gamma$ stands alternatively for $\{cc,cs,ss\}$. This decomposition is 
nicely verified by our numerical date as shown in Fig.\ref{sec:TYPE_CONTACT:ksi_qp}.

\begin{figure}
\includegraphics[width=8cm]{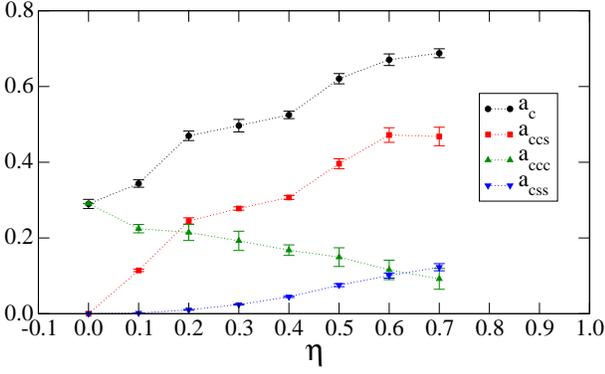}
\caption{Partial contact orientation anisotropies $a_{ccc}$, $a_{ccs}$ and $a_{css}$ 
of $cc$, $cs$ and $ss$ contacts  as the 
function of $\eta$ in the residual state. The error bars represent the standard deviation 
in the residual state.
\label{sec:TYPE_CONTACT:a}}
\end{figure}

Since the contact orientation anisotropy expressed in $(\bm n,\bm t)$ frame 
and the force anisotropy expressed in  $(\bm {n'},\bm{t'})$ frame 
provide respectively fine descriptions of the geometrical and force 
organizations (see section \ref{Geom_mec_origins}), we restrict here 
our analysis to the contribution of various contact types to 
the contact orientation anisotropy $a_c$ and the radial force anisotropies $a_{fn'}$ and 
$a_{ft'}$. 
Figure \ref{sec:TYPE_CONTACT:a}  shows the variation of the partial contact 
anisotropies $a_{ccc}$, $a_{ccs}$ and $a_{css}$ due to  
$cc$, $cs$ and $ss$ contacts in the residual state
as the function of $\eta$. 
The anisotropy $a_{css}$ of  $ss$ contacts increases
slowly with $\eta$ from $0$ to $0.18$. At the same time, 
$a_{ccc}$ decreases and at $\eta=0.7$ we have $a_{css}=a_{ccc}$.   
Hence, although the $ss$ contacts 
represent at $\eta=0.7$ nearly $20\%$  of contacts, their contribution 
to the contact anisotropy remains modest and of the same order as $cc$ contacts.  
The variation of the contact anisotropy $a_c$ is thus largely  
governed by that of $a_{ccs}$.  

\begin{figure}
\includegraphics[width=8cm]{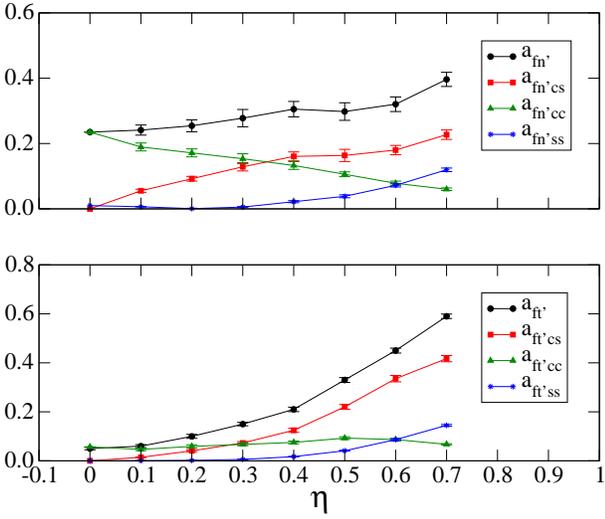}
\caption{Partial radial force anisotropies $a_{fn'cc}$, $a_{fn'cs}$ and $a_{fn'ss}$, and 
 partial orthoradial force anisotropies
$a_{ft'cc}$, $a_{ft'cs}$ and $a_{ft'ss}$ for different contact types  as a function of
$\eta$ in the residual state. The error bars represent the standard deviation 
in the residual state.
\label{sec:TYPE_CONTACT:an}}
\end{figure}

Figure \ref{sec:TYPE_CONTACT:an} shows the partial radial force anisotropies 
$a_{fn'cc}$, $a_{fn'cs}$ and $a_{fn'ss}$, as well as the partial 
orthoradial force anisotropies
$a_{ft'cc}$, $a_{ft'cs}$ and $a_{ft'ss}$ in the residual state 
as the function of $\eta$. As for contact anisotropies, 
the $cs$ contacts carry most of the radial and orthoradial
force anisotropies. The $ss$ contacts contribute modestly to the global force anisotropies 
only for $\eta \geq 0.4$. The anisotropy  $a_{fn'cc}$
declines with $\eta$, mainly due to their low number, and $a_{ft'cc}$ 
stays nearly constant.

\begin{figure}
\includegraphics[width=9cm]{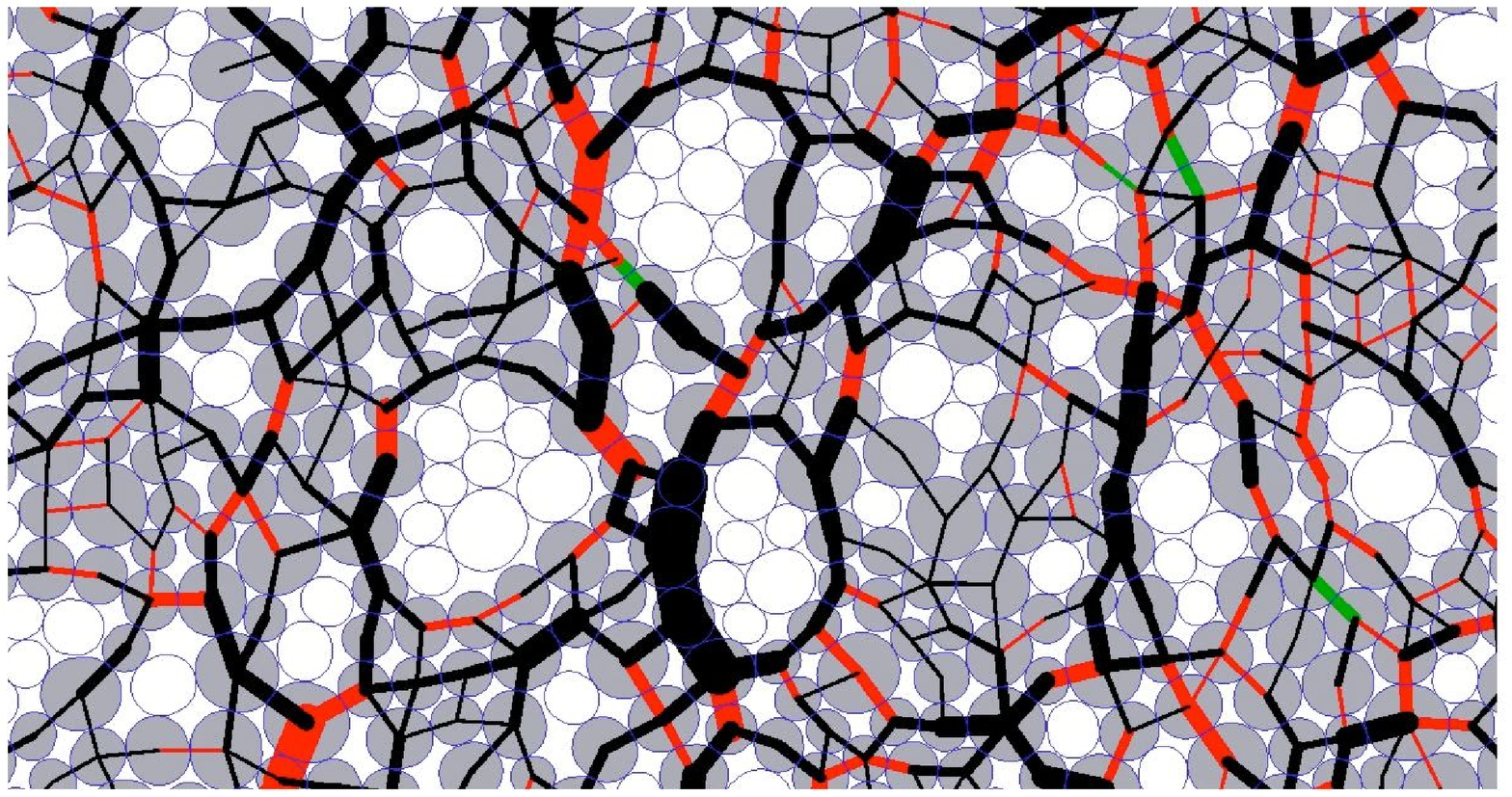}
\includegraphics[width=9cm]{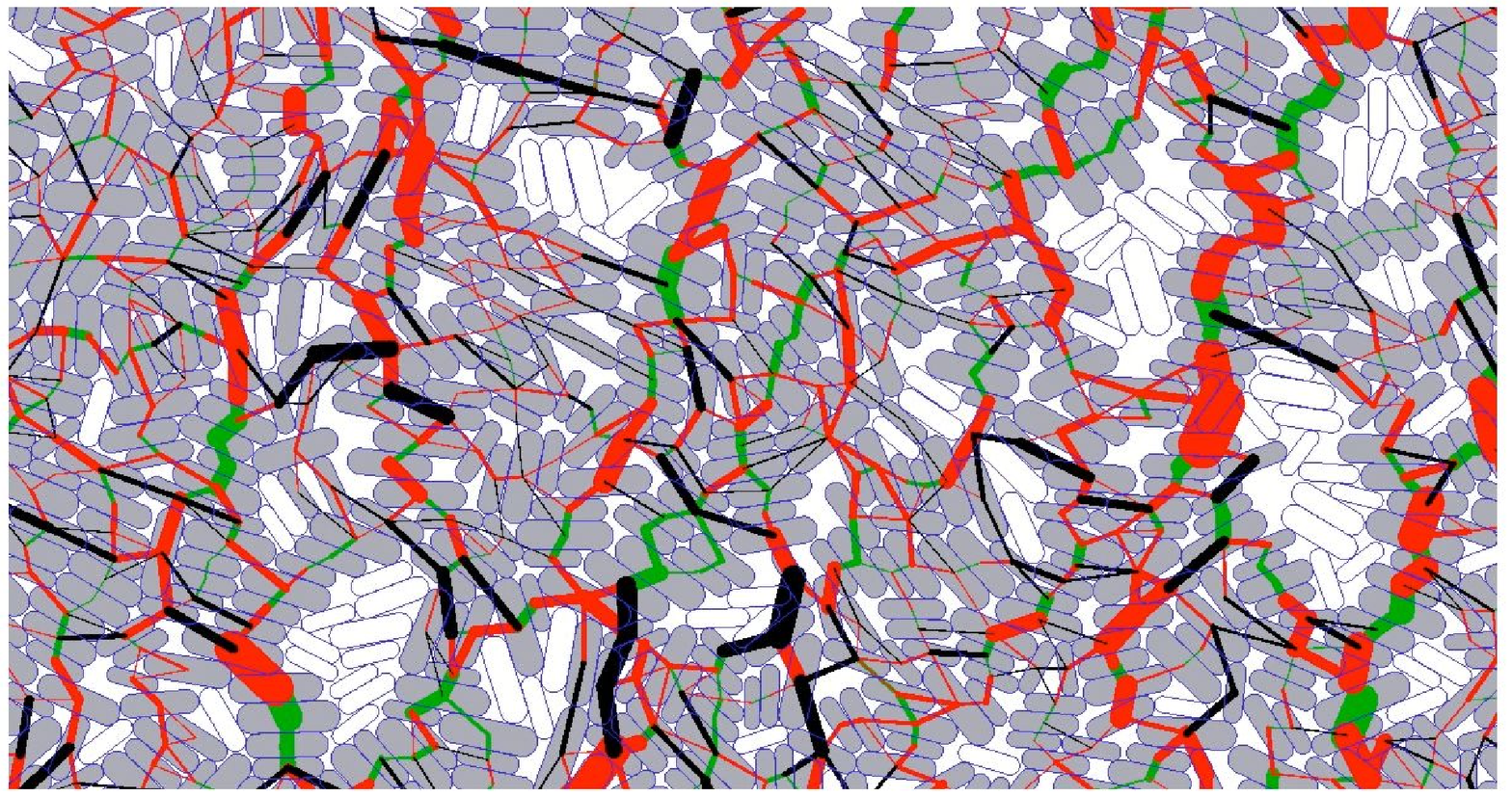}
\caption{(Color online) Snapshot of radiale forces for $\eta=0.2$ (up) and $\eta=0.7$ (down). Line thickness is proportional to the radial 
force. The cap-to-cap, cap-to-side and side-to-side contacts are in black, in red (dark gray)
and in green (light gray).
\label{sec:TYPE_CONTACT:map}}
\end{figure}

A map of contact forces projected 
along the branch vectors is displayed in Fig. \ref{sec:TYPE_CONTACT:map} 
in different colors according to the type of contact.
For $\eta=0.7$, we see that the network of very strong zigzag force 
chains is composed mostly 
of $cs$ and $ss$-contacts and occasionally mediated by $cc$ contacts. 
In contrast, for $\eta=0.2$, the $cc$ contacts appear 
clearly to be correlated in the form of long  chains across the packing 
rarely mediated by $cs$-contacts. 
In all cases, the strong force chains are mostly parallel to the direction of compression.

\begin{figure}
\includegraphics[width=8cm]{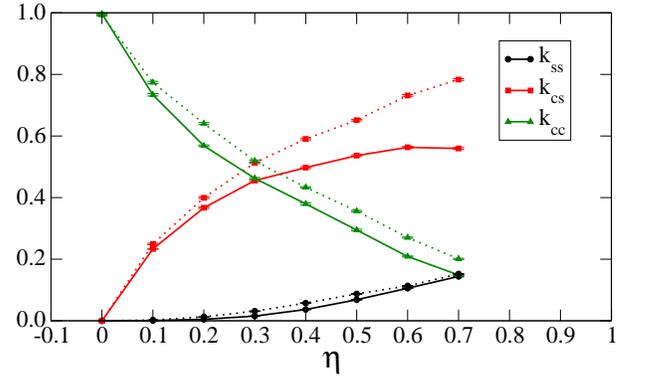}
\caption{Proportions of cap-to-cap ($k_{cc}$), cap-to-side ($k_{cs}$) and side-to-side ($k_{ss}$) contacts
in the strong (plain line) and week (dashed line) networks as a function of $\eta$ in 
the residual state.  The error bars represent the standard deviation 
in the residual state.
\label{sec:TYPE_CONTACT:strong}}
\end{figure}

In order to recognize quantitatively the roles of $cc$, $cs$ and $ss$ 
contacts with respect 
to the force network, we plot in Fig. \ref{sec:TYPE_CONTACT:strong}  
their respective proportions 
$k_{cc}$, $k_{cs}$ and $k_{ss}$  alternatively for the strong and 
weak networks in the residual state as 
as a function  $\eta$. Notice that the data are normalized for each network. 
The proportion of $cc$ contacts declines rapidly in both networks as 
$\eta$ increases whereas that of $cs$ and $ss$ contacts grow. 
We also remark that the  $cs$ contacts are slightly more 
numerous in the weak network than in the strong network. 
The proportions have nearly the same value in the two networks for 
$cc$ and $ss$ contacts.   

\section{Conclusions}
\label{conclusion}

In this paper, we investigated the effect of particle elongation on  
the quasistatic behavior of sheared granular materials by means of  Contact Dynamics 
simulations. The particle shapes are rounded-cap rectangles (RCR) characterized by 
their elongation $\eta$ defined as deviation 
from a reference circular shape, or alternatively by their aspect ratio. 
As the elongation  increases from 0 to 1, the particle shape  
varies continuously from a disk to an increasingly thin rectangle with rounded caps.     
The macroscopic and microstructural properties of several packings 
of 13000 particles, subjected to biaxial compression, were analyzed 
as a function of $\eta$. 

An interesting finding is that the shear strength is an increasing linear function 
of elongation, suggesting that the  parameter $\eta$ is a ``good'' shape parameter
for our 2D granular packings. In order to understand the microscopic origins of
this behavior, we performed an additive decomposition of the stress tensor 
based on a harmonic approximation of the angular variation of average local 
branch vectors, contact normals and forces. This approximation of the shear 
strength in terms of texture and force anisotropies turns out to be in excellent
agreement with our numerical data in the investigated range of the elongation
parameter ($\eta \in[0,0.7]$). Given the evolution of various anisotropies  
with particle elongation, we find that both force and texture anisotropies 
contribute to the increase of 
shear strength, but the increasing mobilization of friction force and the associated
anisotropy seem to be the key effect of particle elongation. In particular 
the proportion of sliding contacts increases strongly as the particles become
more elongated. This effect is correlated with a local nematic ordering of the 
particles which tend to be oriented perpendicular to major principal stress direction.
This ordering is enhanced beyond $\eta=0.4$ but remains essentially of local
nature. In this respect, the fraction of side-to-side contacts increases at large particle 
elongations. However, the force transmission is found to be mainly guided by 
cap-to-side contacts, 
which represent the largest fraction of contacts for the most elongated particles
and carry a large part of the shear strength.

In contrast to shear strength, the solid fraction is not a monotonous function of 
particle elongation; It first increases with particle elongation, then declines as 
the particles become more elongated. In other words, small deviation from 
circular shape favors the space-filling aptitude of the particles but beyond
a characteristic elongation the excluded-volume effects prevail and lead 
to increasingly larger pores which cannot be filled by the particles.
It is remarkable that the coordination number does not follow the solid fraction
but increases with particle elongation, so that the packings of the most elongated
particles are loose but well connected. 

Some features discussed in this paper can legitimately be attributed to the 
two-dimensional
geometry of the particles. For example, rounded-cap-cylinders (sphero-cylinders), 
as three-dimensional analogs of RCR particles in 2D, do not undergo 
spontaneously a nematic ordering.  However, we expect that 3D packings of 
rounded-cap-cylinders behave in many ways as our 2D packings with 
increasing particle elongation. In particular, the excluded volume effect is reinforced by particle elongation and it leads to a similar nonmonotonous dependance on the elongation as in 2D 
\cite{Donev2004,Donev2004a,Donev2007}. In any case, it would be
highly instructive to investigate 3D packings
of elongated particles, along the same lines as in this paper.
It is also obvious that more work required to assess the proper role of friction 
in 2D for RCR particles since friction mobilization seems to underlie to a large
extent the shear strength, we are  performing presently more simulations
 with lower values of friction coefficient. In the same way, we consider the
 effect of cohesion between particles with respect to the shear strength of packings 
 of elongated particles.

We specially thank I. Zuriguel and F. Dubois for fruitful discussions. 
This work was done as part of PPF CEGEO research project 
(www.granulo-science.org/CEGEO).

\bibliography{./azema}

\end{document}